\documentclass[reprint,showkeys, amsmath,amssymb, aps, floatfix]{revtex4-2}

\usepackage{graphicx,subfigure}
\graphicspath{{figures/}}
\usepackage{dcolumn}
\usepackage[outdir=./epsTopdf/]{epstopdf}
\usepackage{bm}
\usepackage{amsmath}
\usepackage{amssymb}
\usepackage{physics}
\usepackage{xcolor}
\usepackage{slashed}
\usepackage{multirow}
\usepackage{float}
\usepackage{caption}
\usepackage{comment}
\usepackage{ mathrsfs }
\usepackage[colorlinks=true,linkcolor=blue,urlcolor=blue,citecolor=blue]{hyperref}

\begin{document}
	
	\preprint{APS/123-QED}
	
	\title{Role of polarizations and spin-spin correlations of $W's$ in $e^-e^+ \to W^-W^+$ \\ at  $\sqrt{s} = 250$ GeV to probe anomalous $W^-W^+Z/\gamma$ couplings}
	
\author{Amir Subba}
\email{as19rs008@iiserkol.ac.in}

\author{Ritesh K. Singh}
\email{ritesh.singh@iiserkol.ac.in}
\affiliation{Department of Physical Sciences, Indian Institute of Science Education and Research Kolkata, Mohanpur, 741246, India\\
	}
	
\date{\today}

\begin{abstract}
We study anomalous $W^-W^+Z/\gamma$ couplings due to dimension-$6$ operators 
in the production process $e^-e^+ \to W^-W^+$ followed by semi-leptonic decay
using polarizations and spin-spin correlations of $W$ bosons. The construction 
of some of the polarization and spin-spin correlation asymmetries required one 
to distinguish between two decay quarks coming from $W^+$ decay. We developed
an artificial neural network (ANN) and a boosted decision tree (BDT) to
distinguish down-type jets from up-type jets and used them to put constraint
on anomalous couplings at International Linear Collider (ILC) running at 
$\sqrt{s} = 250$ GeV with integrated luminosities of $\mathcal{L} \in
\{100 \ \text{fb}^{-1}, \ 250 \ \text{fb}^{-1}, \ 1000 \ \text{fb}^{-1}, \ 
3000 \ \text{fb}^{-1}\}$.
We find that the use of polarization and spin correlation observables, on top of
the cross-sections, significantly improves the limits on anomalous coupling 
compared to the earlier studies.
\end{abstract}
	
\maketitle


\section{Introduction}
The $SU(3)_c\times SU(2)_L \times U(1)_Y$ group structure of Standard Model~(SM) predicts a self-interactions of weak gauge bosons. The predicted self interactions i.e. triple and quartic gauge-boson couplings provides a unique testing ground for new fundamental interactions. The couplings related to fermions with gauge bosons predicted by SM is experimentally confirmed by various experiments to high accuracy. With the discovery of SM like Higgs boson at LHC~\cite{CMS:2012qbp,ATLAS:2012yve}, the particle spectrum of SM is complete. Though SM remains the best tested theory for the particle and their interactions till date, yet we have a growing plethora of phenomena that remain unexplained in the domain of SM. It is known ~\cite{Planck:2013pxb} that nearly 80$\%$ of matter of our Universe is dark matter and till now the detailed structure of the dark matter is still a mystery. The recently reported mass of $W$ boson~\cite{CDF:2022hxs} and the magnetic moment of muon~\cite{Muong-2:2021ojo} are in tension with the predictions of SM. All these results along with the theoretical naturalness in the mass of Higgs boson do tell us that the SM is incomplete and the fundamental theory is still out there lurking in the dark. However, experiments have failed to produce any 
significant evidence for the many explicit models of physics beyond the SM~(BSM) 
viz. supersymmetry, models with universal extra dimensions~(UED), Technicolor 
and so on. As a result, one move to a model-independent way to search for a wide
range of possible BSM effects. We follow a model-independent way of expanding SM 
called as effective field theory~(EFT). In this approach, SM is extended by non-renormalizable gague-invariant operators with mass dimensions $D>4$, which encodes the effects of new particles with the mass scale $\Lambda$  much larger than the $W$ boson mass $m_W$. All the higher dimensional operators are constructed out
of the SM fields assuming the new physics is too
heavy that we can integrate them out of the Lagrangian. Assuming lepton-number
conservation, the effective Lagrangian is written as~\cite{Buchmuller:1985jz}
\begin{equation}
\mathscr{L}_{EFT} = \mathscr{L}_{SM} +
\frac{1}{\Lambda^{2}}\sum_ic^{(6)}_i\mathscr{O}^{(6)}_i +
\frac{1}{\Lambda^4}\sum_jc^{(8)}_j\mathscr{O}^{(8)}_j + ..
\end{equation} 
where $c_i^{(6,8)}$ are the Wilson's Coefficient or the couplings of the higher
dimension operators. The effects of the new physics are translated to the weak scale via these Wilson coefficients. In this paper, we study dim-6 effective operators
which gives the anomalous triple gauge couplings~($WWV, V \in Z,\gamma$) and
constrain those couplings~($c^{(6)}_i$). Considering both $CP$-even and odd,
the relevant dim-6 effective operators in the HISZ basis contributing to the $WWV$ couplings
are~\cite{PhysRevD.48.2182,Degrande:2012wf}
\begin{equation}
\begin{aligned}
&\mathscr{O}_{WWW} &=& \quad\text{Tr}[W_{\nu \rho}W^{\mu \nu} W_{\rho}^{\mu}]\\
&\mathscr{O}_{W} &=& \quad(D_\mu \Phi)^\dagger W^{\mu \nu} (D_\nu \Phi) \\
&\mathscr{O}_B &=& \quad(D_\mu \Phi)^\dagger B^{\mu \nu} (D_\nu \Phi) \\
&\mathscr{O}_{\tilde{WWW}} &=& \quad\text{Tr}[W_{\mu \nu} W^{\nu \rho} W^\mu_\rho] \\
&\mathscr{O}_{\tilde{W}} &=& \quad (D_\mu \Phi)^\dagger \tilde{W}^{\mu \nu} (D_\nu \Phi)
\end{aligned}
\label{eft:ope}
\end{equation}
where $\Phi = \begin{pmatrix}
\phi^+ \\ \phi^0
\end{pmatrix}$ is the Higgs double field and $W^{\mu \nu}, B^{\mu \nu}$ represents the full(non-abelian) field strengths of $W$ and $B$ gauge fields and are defined as:
\begin{flalign*}
D_\mu &= \partial_\mu  +\frac{i}{2}g\tau^i W_\mu^i + \frac{i}{2}g'B_\mu \\
W_{\mu \nu} &= \frac{i}{2}g\tau^i(\partial_\mu W_\nu^i-\partial_\nu W_\mu^i + g\epsilon_{ijk}W_\mu^i W_\nu^k) \\
B_{\mu \nu} &= \frac{i}{2}g'(\partial_\mu B_\nu - \partial_\nu B_\mu)
\end{flalign*}
The first three operator are $C$ and $P$ conserving and last two violate $C$ and$/$or $P$. All these operators of Eq.~(\ref{eft:ope}) after electroweak symmetry breaking~(EWSB) give rise to non-standard triple gauge couplings. Conventionally the $WWV$ vertices are parametrized by the effective Lagrangian~\cite{Hagiwara:1986vm,PhysRevD.48.2182}  
\begin{equation}
\begin{split}
\mathscr{L}_{eff}^{WWV} = ig_{WWV}[g_1^V(W^+_{\mu \nu}W^{-\mu} - W^{+\mu}W^-_{\mu \nu})V^\nu  \\
+  k_V W^+_\mu W^-_\nu V^{\mu \nu} +\frac{\lambda_V}{m_W^2}W_\mu^{\nu+}W_\nu^{-\rho}V_{\rho}^{\mu}  \\
+ ig_4^VW_\mu^+W_\nu^-(\partial^\mu V^\nu+\partial^\nu V^\mu) \\ 
- ig_5^V\epsilon^{\mu \nu \rho \sigma}(W_\mu^+ \partial_\rho W_\nu^- - \partial_\rho W_\mu^+W_\nu^-)V_\sigma \\ 
+ \tilde{k}_VW_\mu^+W_\nu^-\tilde{V}^{\mu \nu} + \frac{\tilde{\lambda}_V}{m_W^2}W_\mu^{\nu+}W_\nu^{-\rho}\tilde{V}_\rho^{\mu}]
\end{split} 
\label{Lag:eff}	
\end{equation}
with $g_{WW\gamma} = -e$ and $g_{WWZ} = -e \ \text{cot}\theta_W$. The first three terms of Eq.~(\ref{Lag:eff}) respect $C$ and $P$ and the remaining four terms violate $C$ and$/$or $P$. Within the SM, the couplings are given by $g_1^V=k_V=1$ and other couplings are zero. While the value of $g_1^\gamma,g_4^\gamma,g_5^\gamma$ are fixed by the electromagnetic gauge invariance, the presence of the operators $\mathscr{O}_{WWW},\mathscr{O}_W,\mathscr{O}_B,\mathscr{O}_{\tilde{W}},\mathscr{O}_{\tilde{WWW}}$ in the effective Lagrangian will change the other values to~\cite{Hagiwara:1986vm}
\begin{equation}
\begin{aligned}
g_1^Z &= 1 + c_W\frac{m_Z^2}{2\Lambda^2}\\
k_Z &= 1 + [c_W-s_W^2(c_B+c_W)]\frac{m_Z^2}{2\Lambda^2}\\
k_\gamma &= 1 + (c_B+c_W)\frac{m_W^2}{2\Lambda^2} \\
\lambda_\gamma &= \lambda_Z = c_{WWW}\frac{3m_W^2g^2}{2\Lambda^2}\\
g_4^Z&=g_5^Z=0\\
\tilde{k}_Z &= -c_{\widetilde{W}}s_W^2\frac{m_Z^2}{2\Lambda^2}\\
\tilde{k}_\gamma &= c_{\widetilde{W}}\frac{m_W^2}{2\Lambda^2}\\
\tilde{\lambda}_\gamma &= \tilde{\lambda}_Z = c_{\widetilde{WWW}}\frac{3m_W^2g^2}{2\Lambda^2}
\end{aligned}
\end{equation}
with $s_W = \text{sin}\theta_W$.\\	
The anomalous $WWZ/\gamma$ vertex has been studied extensively at $e^-e^+$~\cite{Zhang:2016zsp,OPAL:2000wbs,OPAL:2003xqq,DELPHI:2008uqu,Rahaman:2019mnz,ALEPH:2013dgf,PhysRevD.30.1881,Gaemers:1978hg,HAGIWARA1992353,Choudhury:1996ni,Choudhury:1999fz,Wells:2015eba,Buchalla:2013wpa,Berthier:2016tkq,Bian:2015zha,Bian:2016umx,Beyer:2020eas}, Large Hadron collider~\cite{Baglio:2019uty,Bian:2015zha,Choudhury:2022iqz,CMS:2021icx,ATLAS:2021jgw,CMS:2021foa,CMS:2021lix,Baur:1987mt,Dixon:1999di,Falkowski:2016cxu,Butter:2016cvz,Azatov:2017kzw,Baglio:2017bfe,Li:2017esm,PhysRevD.96.073003,Bhatia:2018ndx,Chiesa:2018lcs,Rahaman:2019lab,CMS:2016qth,ATLAS:2016bkj,ATLAS:2016zwm,CMS:2013ant,RebelloTeles:2013kdy,ATLAS:2012mec,CMS:2012wlr,ATLAS:2013way,CMS:2013ryd,ATLAS:2017pbb,CMS:2017egm,ATLAS:2017luz,CMS:2017dmo,CMS:2019ppl,CMS:2019nep,CMS:2019efc,Yap:2020xjr,Tizchang:2020tqs,Campanario:2020xaf,Ciulli:2020ygo,CMS:2020olm}, Large Hadron electron collider~(LHeC)~\cite{Biswal:2014oaa,Cakir:2014swa,Li:2017kfk,Koksal:2019oqt,Gutierrez-Rodriguez:2019hek} and Tevatron~\cite{CDF:2007aqs,CDF:2012mnr,D0:2010jca,Krop:2010zza,Mastrandrea:2010nx,D0:2012vhk,D0:2015kta}. We list down the best constrained values for various anomalous couplings~($c_i$) obtained experimentally in Table~\ref{tab:constraint} accordingly.  
\begin{table}[h!]
\caption{The list of tightest constraints observed on the effective operators in $SU(2) \times U(1)$ gauge at 95$\%$ C.L. from various experiments.}
\centering
\begin{ruledtabular}
\begin{tabular}{ccc} 
$c_i^{\mathscr{O}}$&Limits(TeV$^{-2}$)&Remarks \\ \hline 
$c_{WWW}/{\Lambda^2}$&[-0.90,+0.91]&CMS~\cite{CMS:2021foa} \\
$c_W/{\Lambda^2}$&[-2.5,+0.3]&CMS\cite{CMS:2021icx}\\ 
$c_B/{\Lambda^2}$&[-8.78,+8.54]&CMS\cite{CMS:2019ppl}\\ 
$c_{\widetilde{W}}/\Lambda^2$&[-20.0,+20.0]&CMS~\cite{CMS:2021foa}\\ 
$c_{\widetilde{WWW}}/\Lambda^2$&[-0.45,+0.45]&CMS~\cite{CMS:2021foa} 
\end{tabular}
\end{ruledtabular}
\label{tab:constraint}
\end{table} \\
The limits given in the above Table~\ref{tab:constraint} are obtained by varying one parameter at a time and others are kept at zero~(SM value). It has been shown in Ref.~\cite{Rahaman:2019mnz} that using the polarised beam in $e^-e^+$ collider, some of the anomalous couplings are constrained better. In our current article, we construct spin-related like polarization asymmetries and spin-spin correlation asymmetries
alongwith cross-section to constrain the above discussed
anomalous couplings. The use of asymmetries will bring the directional limits on various couplings resulting to better constrain on t hose couplings.\\ \\ We probe $W^-W^+$ production process in $e^-e^+$ collider at $\sqrt{s} =
250$ GeV using unpolarised beams and the $W's$ are decayed semi-leptonically
such that the whole process is defined as 
\begin{equation}
e^- + e^+ \to W^-W^+ \to l^-\nu_ljj
\end{equation} 
Here $l^- \in (e^-,\mu^-)$ and $j's$ are the light quarks viz. $u\bar{d}/c\bar{s}$. The production process $e^-e^+ \to W^-W^+$ proceeds through one neutrino mediated $t$-channel and two $\gamma^*/Z$ mediated $s$-channel. The $s$-channel diagrams contains trilinear $WWV$ gauge boson couplings whose deviations from SM value in presence of Lagrangian given by Eq.~(\ref{Lag:eff}) is studied in this article. We perform our analysis at the particle level
i.e. the quarks obtained at the matrix element level are allowed to undergo showering and hadronization.
While constructing some the spin-related asymmetries, we need the correct information of daughter particles of $W$ boson and for that we used machine learning~(ML) techniques particularly artificial neural network~(ANN) and
boosted decision trees~(BDT).  \\ \\
We describe in Section~\ref{spinobs} spin and the observables obtained
using spin of a particle. We mostly focused on the asymmetries and the spin-spin
correlation asymmetries of spin-1 boson. We also list down the relevant
observables affected by flavor tagging and those which are not. A method of ML
techniques used for flavour tagging the jets to the light quarks is explained in
section~\ref{ml}. In section~\ref{Pest}, we discuss parameter estimation and the
limits obtained on the five anomalous couplings. We conclude in
section~\ref{conclude}.
		\section{Spin and related Observable}
	\label{spinobs}
	All the fundamental particles have finite spin and all the fundamental
	interactions conserve angular momentum. The spin of a given particle decides the
	Lorentz structure of the couplings it will have with other particles and hence
	its production and decay mechanism. The spin and polarization information of a
	decaying particle gets encoded in the angular distribution of its decay
	products. One can use various kinematical distributions of decay products to
	decode the spin content and dynamics of any process~\cite{leader_2001}.
	The range of spin and
	polarization sensitive observables that we discuss in this section are broadly
	divided in two class: The first one deals with asymmetries that measure various
	polarization parameters of a resonance. The second class involves asymmetries
	that probe the spin-spin correlation between two resonances. We discuss them
	here one by one.
	\subsection{Polarization asymmetries}
\begin{figure}[h!]
		\includegraphics[scale=0.5]{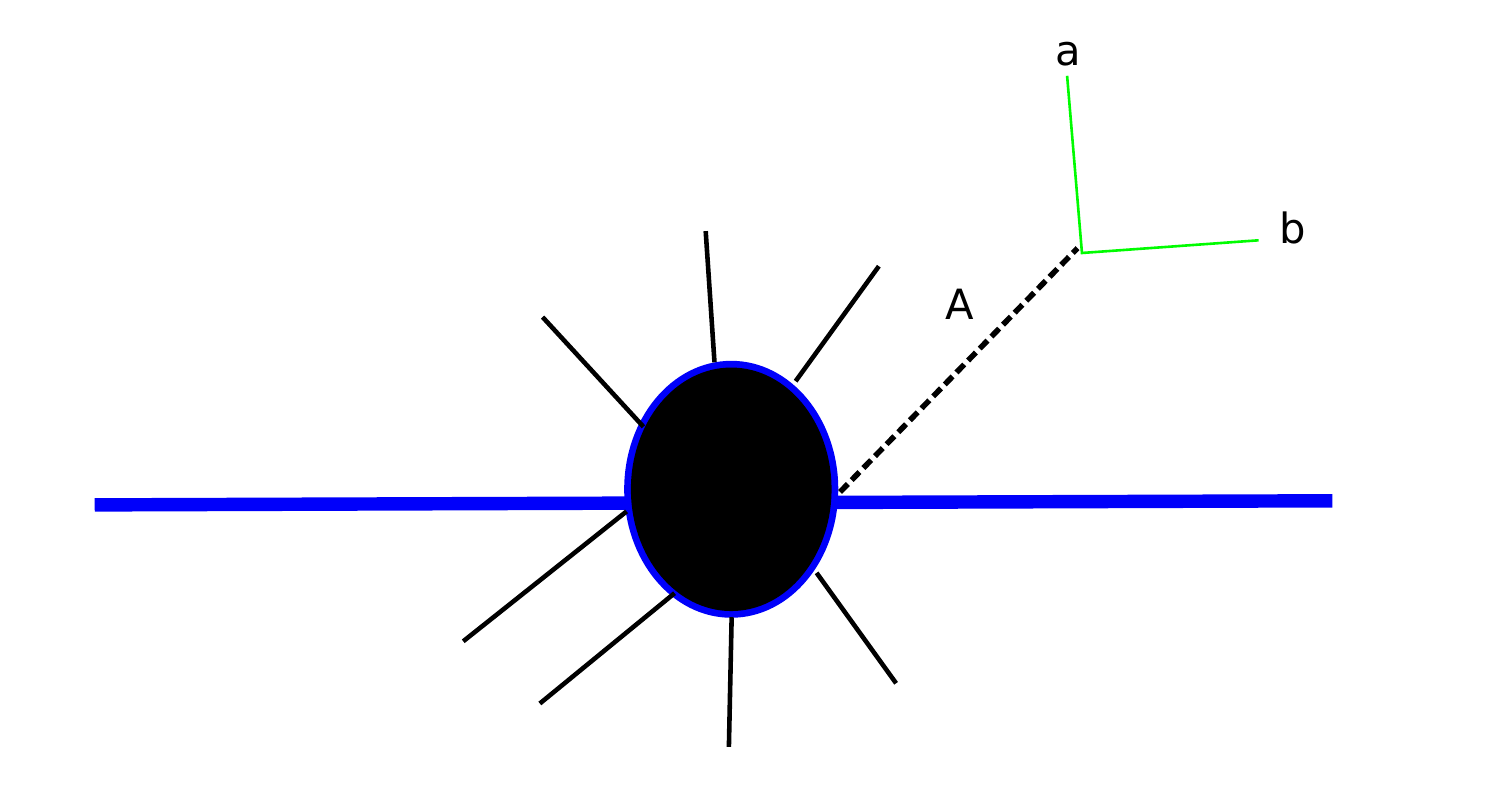}
		\caption{Schematic diagram showing the production of a resonance $A$ in a
			scattering process followed by its decay to particles $a$ and $b$.}
		\label{figure:1pdecay}
	\end{figure}

	Let us consider a scattering process of particle $B_1$ and $B_2$
	where a resonance $A$ of spin $s$ along with some other particles is produced
	followed by its subsequent decay to $a$ and $b$, shown in
	Fig.~\ref{figure:1pdecay}.
	The differential rate for such process, assuming narrow width 
	approximation~(NWA) for resonance $A$, is given as~\cite{Boudjema:2009fz},
	\begin{equation}
		\begin{split}
			d\sigma &= \sum_{\lambda,\lambda'}\left[
			\frac{(2\pi)^4}{2I_{B_1B_2}}\rho(\lambda,\lambda')\delta^4\left(k_{B_1}+k_{B_2}-P_A-\sum
			p_i\right) \right.\\ & \left. \frac{d^3p_A}{2E_A(2\pi)^3}\prod_i
			\frac{d^3p_i}{2E_i(2\pi)^3} \right]
			\times\left[\frac{1}{\Gamma_a}\frac{(2\pi)^4}{2m_A}
			\Gamma'(\lambda,\lambda')\right.\\
			&\left.\delta^4\left(P_A-P_a-P_b\right)\frac{d^3p_a}{2E_a(2\pi)^3}
			\frac{d^3p_b}{2E_b(2\pi)^3}\right] \ ,
		\end{split}
	\end{equation}
	where $I_{B_1B_2}$ is the flux factor and $\lambda's,\Gamma_A,m_A$ are the
	helicities, total width and mass of $A$ respectively. Rewriting the two terms of
	the differential rate in terms of polarisation density,~$P_A(\lambda,\lambda')$
	and decay density matrix,~$\Gamma_A(\lambda,\lambda')$, the decay angular
	distribution can be written down as
	\begin{equation}
		\frac{1}{\sigma}\frac{d\sigma}{d\Omega_a}= \frac{2s+1}{4\pi}
		\sum_{\lambda,\lambda'}P_A(\lambda,\lambda')\Gamma_A(\lambda,\lambda') \ .
	\end{equation}
	The production dynamics is encoded in the given polarisation density matrix and
	one can calculate to quantify the production rate of various quantum
	interference states. The general expressions for $P_A(\lambda,\lambda')$ and
	$\Gamma_A(\lambda,\lambda')$ for a spin-1 particle are given in 
	Eqs.~(\ref{eqn:A1}) and (\ref{eqn:A2}), respectively. Using these expressions
	one can rewrite the angular distribution for a spin-1 particle as
	\begin{equation}
		\label{eqn:diffn}
		\begin{aligned}
			\frac{1}{\sigma}\frac{d\sigma}{d\Omega_a} &=
			\frac{3}{8\pi}[(\frac{2}{3}-(1-3\delta)\frac{T_{zz}}{\sqrt{6}}) + \alpha p_z
			\text{cos}\theta_a \\ &+ \sqrt{\frac{3}{2}}(1-3\delta)T_{zz}\text{cos}^2\theta_a
			\\ & + (\alpha p_x +
			2\sqrt{\frac{2}{3}}(1-3\delta)T_{xz}\text{cos}\theta_a)\text{sin}\theta_a
			\text{cos}\phi_a \\ &+ (\alpha p_y +
			2\sqrt{\frac{2}{3}}(1-3\delta)T_{yz}\text{cos}\theta_a \text{sin}\theta_a
			\text{cos}\phi_a \\ & + (1-3\delta)(\frac{T_{xx} -
				T_{yy}}{\sqrt{6}})\text{sin}^2\theta_a \text{cos}(2\phi_a) \\ &+
			\sqrt{\frac{2}{3}}(1-3\delta)T_{xy}\text{sin}^2\theta_a \text{sin}(2\phi_a)], 
		\end{aligned}
	\end{equation}
	where $\theta_a, \phi_a$ are the polar and azimuthal angle of daughter $a$  
	in the rest frame of parent $A$ with its would be
	momentum along the $z$-axis. The initial beam direction and the $A$ momentum
	in the lab frame define the $x$-$z$ plane, i.e. $\phi=0$ plane, in the rest frame
	of $A$ as well. For $A$ being a vecor boson decaying to a pair of fermion
	through $V$-$A$ interaction, the parameters $\alpha$ and $\delta$ 
	are given by~\cite{Boudjema:2009fz},
	\begin{equation}
		\begin{aligned}
			\alpha &= \frac{2(R^2_a-L^2_a)\sqrt{1+(x^2_1 - x^2_2)^2 - 2(x^2_1+x^2_2)}}
			{12L_a R_a x_1x_2+C(R_a^2+L_a^2)^2} \ ,\\
			\delta &= \frac{4L_a R_a x_1x_2 + (R_a^2+L_a^2)(C-2)}{12R_a L_ax_1x_2+C(R_a^2+
				L_a^2)} \ ,
		\end{aligned}
	\end{equation} 
	where $C = 2-(x_1^2-x_2^2)^2+(x_1^2+x_2^2), x_i = \frac{m_i}{M_A}$. In the high energy limit the final state
	fermions ($e^\mp, \mu^\mp, u, d , c, s$) can be taken to be massless which
	implies $x_1 \to 0, x_2 \to 0$ and $\delta \to 0, \alpha \to
	\frac{R^2_a-L^2_a}{R^2_a+L^2_a}$. Further, for the decay of $W$, within the
	$SM$, $R_a = 0$ hence $\alpha = -1$. The vector~$\Vec{p}$ and tensor~$T_{ij}$
	polarization can be calculated from the production part. For example $P_x$ and
	$T_{xz}$ can be calculated as follows:
	\begin{equation}
		\begin{aligned}
			\label{eq:A_rho}
			P_x &= \frac{[[\rho_T(+,0)+\rho_T(0,+)]+[\rho_T(0,-)+\rho_T(-,0)]]}{\sqrt{2}
				\sigma} \ ,\\ 
			T_{xz} &= \frac{\sqrt{3}[\rho_T(+,0)+\rho_T(0,+)]-[\rho_T(0,-)+\rho_T(-,0)]}{4
				\sigma} \ .
		\end{aligned}
	\end{equation}
	All other polarizations can be found from the different combination of density
	matrix and tracelessness of $T_{ij}$ as shown in \cite{Rahaman:2017qql,Rahaman:2021fcz}.
	
	\begin{table}[tb!]
		\caption{Table showing the asymmetries of a spin-1 particle and the angular parameters that are used to find the respective asymmetries}
		\begin{ruledtabular}
					\begin{tabular}{@{}p{1.5cm}@{}p{2.5cm}@{}p{4.0cm}@{}} 
						$A_i$ & $c_j$ & functions \\ \hline
						$A_x$& $c_1 \equiv c_x$ &sin$\theta$cos$\phi$ \\ 
						$A_y$&  $c_2 \equiv c_y$ &sin$\theta$sin$\phi$ \\ 
						$A_z$&  $c_3 \equiv c_z$ &cos$\theta$ \\
						$A_{xy}$&  $c_4 \equiv c_{xy}$ &sin$^2\theta$sin$(2\phi)$ \\
						$A_{xz}$&  $c_5 \equiv c_{xz}$ &sin$\theta$cos$\theta$cos$\phi$ \\ 
						$A_{yz}$&  $c_6 \equiv c_{yz}$ &sin$\theta$cos$\theta$sin$\phi$ \\
						$A_{x^2-y^2}$&  $c_7 \equiv c_{x^2-y^2}$ &sin$^2\theta$cos$(2\phi)$ \\ 
						$A_{zz}$&  $c_8 \equiv c_{zz}$ &sin$(3\theta)$\\ 
					\end{tabular}
				\end{ruledtabular}
				\label{tab:asym_par}
			\end{table}
			Similarly at the level of decay products, one can find the same polarization
			parameters by using different asymmetries constructed from the decay angular
			distribution of fermions. Different asymmetries can be calculated using
			different combination of angular variables ($c_j$) as given below: 
			\begin{equation}
				A_i = \frac{\sigma(c_j > 0)-\sigma(c_j < 0)}{\sigma(c_j > 0)+\sigma(c_j < 0)} \
				.
				\label{eq:A_sig}
			\end{equation}
			Here $c_j$ is a function of $\phi$ and $\theta$ of the final state fermions.
			The relation of different asymmetries with the angular functions $c_j$ is listed
			in Table~\ref{tab:asym_par}.
			The angular functions $c_1$~--~$c_3$ are parity odd while $c_4$~--~$c_8$ and
			parity even. This means that asymmetries $A_1$~--~$A_3$ can be non-zero only if
			there is parity violation in the decay process i.e. $\alpha \neq 0$, as these
			three asymmetries are proportional to the $\alpha$ parameter. Other five
			asymmetries $A_4$~--~$A_8$ are non-zero as long as the corresponding tensor
			polarization appearing in Eq.~(\ref{eqn:diffn}) are non-zero. 
			
			For two body decay in the rest frame, two daughters emerge in opposite
			directions, i.e. if we average over them then the asymmetries $A_1$~--~$A_3$
			will vanish. In other words, to construct the vetor polarizations $\vec{p}$ we
			need to be able to distinguish between two daughters. This is possible in the
			leptonic decays of $W^\pm$ but for the hadronic decay channels we need a method
			to identify (tag) them. This issue is addressed in section~\ref{ml}.
			\subsection{Spin-Spin Correlation}
			\begin{figure}[tb!]
				\includegraphics[scale=0.8]{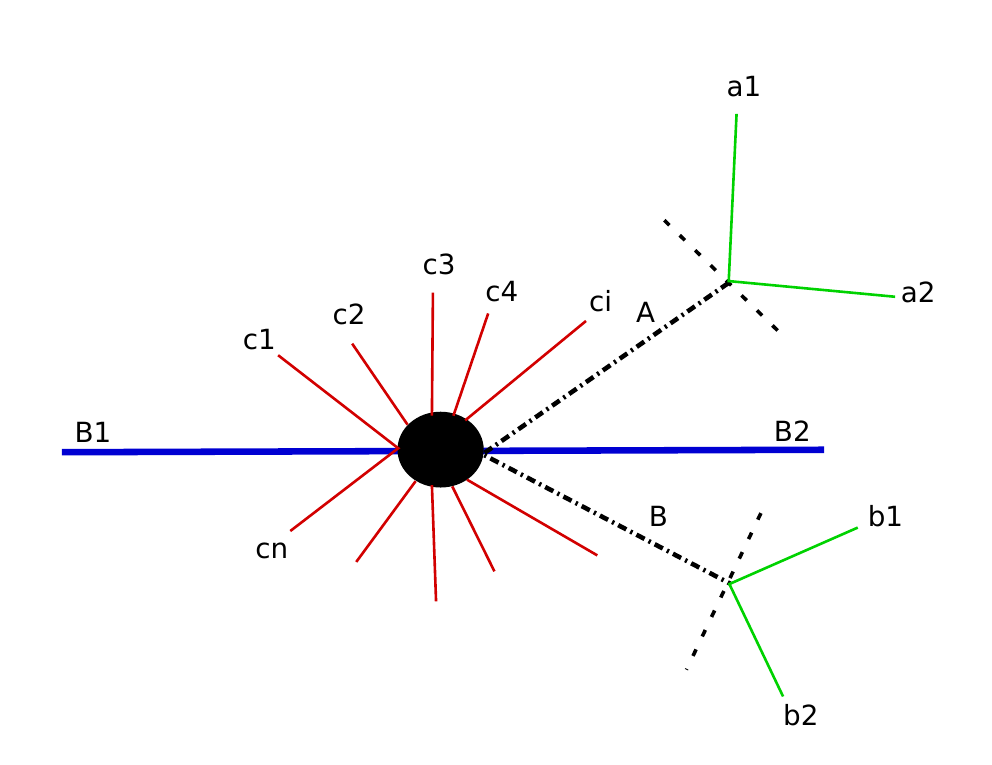}
				\caption{\label{fig:2decay}Schematic diagram showing the production of two
					resonance $A$ and $B$ followed by their decay to $a1a2$ and $b1b2$
					respectively.}
			\end{figure} 
			For a polarization asymmetry to be non-zero we need the corresponding particle
			being produced with non-zero polarization. In the case of unpolarized beam
			collisions producing a pair of fermions, we require parity violation in the
			production process to have a non-zero polarization. But for $t\bar t$ pair
			production at LHC through QCD interactions, we have no parity violation and
			hence unpolarized top quarks. However, due to the vectorial nature of the
			gluon's interaction with top-quark one has a certain kind of spin-spin
			correlations between $t$ and $\bar t$ spins. And 
			experiments~\cite{D0:2015kta,ATLAS:2014abv,ATLAS:2012ao} 
			have shown the spin correlations in $t\bar{t}$ systems. These additional set of
			observables will provide an additional probe for the possible 
			NP signal. The spin-spin correlation asymmetries can be calculated in the
			similar fashion as we have shown for single particle calculations of
			asymmetries. We consider a generic scattering process in which two spin-full
			resonance is produced followed by its decay as shown in Fig.~\ref{fig:2decay}.
			The differential rate for this process would remain the similar to 
			Eq.~(\ref{eqn:diffn}) but with few changes: The single particle density matrix
			$\rho(\lambda,\lambda')$ is replaced with two particle density matrix
			$\rho(\lambda_A,\lambda_A',\lambda_B,\lambda_B')$ and there is an additional 
			factor of square bracket terms containing $\Gamma(\lambda,\lambda)$  one for
			decay of particle $A$ and another for particle $B$.
			The full spin correlated polarization density matrix for a pair of spin-1
			particles defined as
			$$P_{AB}(\lambda_A,\lambda_A^{'},\lambda_B,\lambda_B^{'}) =
			\rho(\lambda_A,\lambda_A',\lambda_B,\lambda_B')/{\rm Tr}(\rho) $$
			can be parametrized in terms of polarizations and spin correlation variables 
			as~\cite{Rahaman:2021fcz}:
			\begin{eqnarray}
				\label{corrpol}
				&~~&P_{AB}(\lambda_A,\lambda_A^{'},\lambda_B,\lambda_B^{'})  =
				\frac{1}{9}[I_{3\times3}\otimes I_{3\times3}\nonumber \\ & +&
				\frac{3}{2}\Vec{p}^A.\Vec{S}\otimes I_{3\times3} +
				\frac{3}{2}I_{3\times3}\otimes \Vec{p}^B.\Vec{S} \nonumber\\  & +&
				\sqrt{\frac{3}{2}}T_{ij}^A(S_iS_j + S_jS_i)\otimes I_{3\times3} \nonumber\\ & +&
				\sqrt{\frac{3}{2}}I_{3\times3}\otimes T_{ij}^B(S_iS_j+S_jS_i)  +
				pp_{ij}^{AB}S_i \otimes S_j \nonumber\\ 
				& +& pT_{ijk}^{AB}S_i \otimes (S_jS_k + S_kS_j) \nonumber\\
				&+ & Tp_{ijk}^{AB}(S_iS_j+S_jS_i)\otimes S_k \nonumber\\ & +&
				TT_{ijkl}^{AB}(S_iS_j+S_jS_i)\otimes (S_kS_l+S_lS_k)], \\ \nonumber&& (i,j,k = x,y,z)
			\end{eqnarray}
			The indices $\lambda_A$ and $\lambda_A^{'}$ label the left matrices in the tensor
			products while $\lambda_B$ and $\lambda_B^{'}$ label the right matrices.
			By combining Eq.~(\ref{corrpol}) and (\ref{eqn:A2}), the normalized joint decay
			angular distribution can be written as:
			\begin{equation}
				\label{jdm}
				\begin{split}
					\frac{1}{\sigma}\frac{d^2\sigma}{d\Omega_a d\Omega_b} =
					\left(\frac{3}{4\pi}\right)^2 \sum_{\lambda's}P_{AB}(\lambda_A,
					\lambda_{A'};\lambda_{B},\lambda_{B'})\\
					\Gamma_A(\lambda_A,\lambda_{A'})  \ \ \Gamma_B(\lambda_{B},\lambda_{B'}),
				\end{split}
			\end{equation}
			where $d\Omega_a$ and $d\Omega_b$ are the solid angle measure for the decay
			product $A$ and $B$ particle. In order to compute the various polarization and
			spin correlation parameters appearing in Eq.~(\ref{corrpol}) we can define
			asymmetries similar to Eq.~(\ref{eq:A_rho}) and/or Eq.~(\ref{eq:A_sig}) as
			described in detail in Ref.~\cite{Rahaman:2021fcz}. For the present work, we 
			choose to discuss the asymmetries defined using angular distribution 
			Eq.~(\ref{jdm}) of the final state decayed particles. The set of spin
			correlation asymmetries can be defined as:
			\begin{equation}
				A_{ij}^{AB} = \frac{\sigma(c_i^ac_j^b > 0) - \sigma(c_i^ac_j^b < 0)}
				{\sigma(c_i^ac_j^b > 0) + \sigma(c_i^ac_j^b < 0)} 
				\label{eq:spincorr}
			\end{equation} 
			where $i,j \in (1..8)$ and the parameters $c's$ can be read out from the 
			previous section. These asymmetries probe the spin correlation parameters like:
			$pp_{ij}^{AB}$, $pT_{ij}^{AB}$, $Tp_{ij}^{AB}$ and $TT_{ij}^{AB}$ which are 
			vector-vector, vector-tensor, tensor-vector and tensor-tensor correlations,
			respectively. For example, $A_{13}^{AB}$ probes $pp_{xz}^{AB}$, $A_{14}^{AB}$
			probes $pT_{xxy}$ correlation parameters, etc.\\
			\begin{figure}[h!]
				\includegraphics[width=0.5\textwidth]{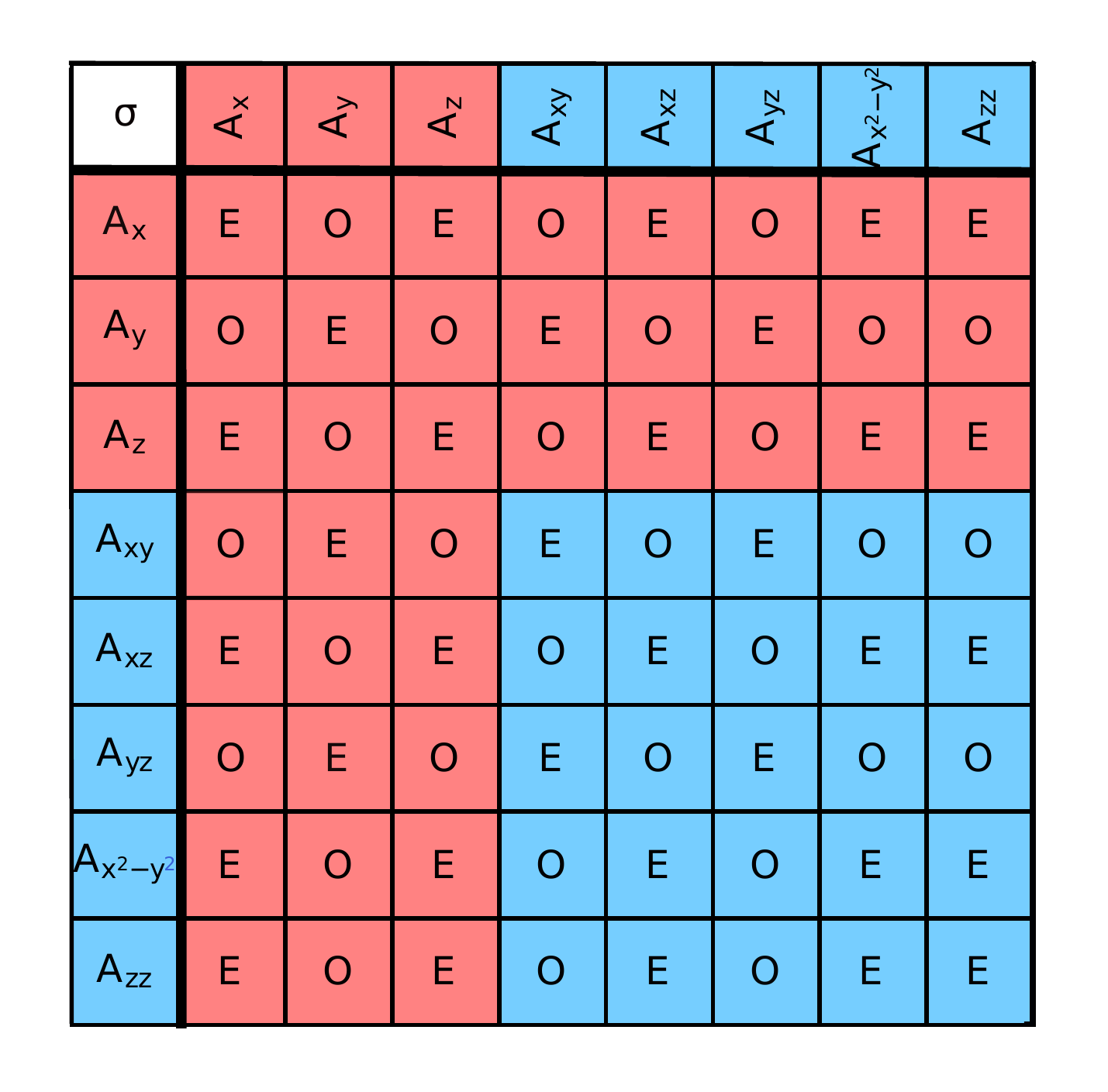}
				\caption{\label{fullcorr}
					Table of polarization asymmetries (first row and first column) and spin
					correlation asymmetries (all others) along with their $CP$ parities. For spin
					correlation asymmetries "E" stands for $CP$-even and "O" for $CP$-odd. The
					polarization asymmetries linear in $y$ are also $CP$-odd and others are
					$CP$-even. The  color {\em light red} indicates that the asymmetries 
					require flavor tagging and {\em ligh blue} indicates immunity to the flavors. 
				}
			\end{figure}
			
			Futhermore, the angular functions $c_1, c_2$ and $c_3$ require the identifiction
			of the flavor of the daughter particles hence some of the spin correlation
			asymmetries will vanish if we average over the flavor in the case of hadronic
			decay of $W^\pm$. Thus for the semi-leptonic final state that we are considering
			in this work, it is utmost important to do a flavor tagging to be able to
			construct all the asymmetrie. Additionally the angular functions $c_2$, $c_4$
			and $c_6$ are $CP$-odd so the corresponding polarization asymmetries are
			expected to be zero for the $CP$ conserving production processes. In this case
			all the spin-correlation having one (and exactly one) factor of $CP$-odd angular
			fuction will also vanish. Examples are $A_{12}$, $A_{34}$ etc. However the
			correlation asymmetries depending upon two of the $CP$-odd angular functions are
			$CP$-even and will be non-zero in general, even in the absence of $CP$-violation
			in the production process. The $CP$ properties of all the polarization and spin
			correlation asymmetries are listed in Fig.~\ref{fullcorr} with letters "E" and
			"O" denoting $CP$-even and -odd, respectively. The asymmetries that require
			flavor reconstructions are marked in {\em light red} color and the asymmetries
			that are immune to the flavor are marked in {\em light blue}. Thre are a total
			of 25 spin correlation asymmetry and 10 polarization asymmetry that are
			independent of flavor tagging, while thre are 39 spin correlation asymmetries
			and 6 polarization asymmetries that will vanish without flavor tagging. The 
			flavor tagging is trivial in the leptonic branch and very obscure in the
			hadronic branch. In the next section, we try to use machine learning methods to
			train models (tagger) to tag the light quark flavors for our purpose in the
			$e^-e^+\to W^-W^+$ process with $W^-$ decaying leptonically and $W^+$ decaying
			hadronically.

			\section{Flavour Tagging}
			\label{ml}
			With the aim to exploit all the asymmetries, we need to develop a
light flavor tagger for $W'$s decaying hadronically and for that we develop
artificial neural network~(ANN) and boosted decision tree~(BDT). These kind of
techniques has long been used in the field of high energy physics. It was used
for track reconstruction in wire chambers~\cite{Neuhaus:2017trg} and cluster
finding in cellular calorimeters~\cite{Denby:1987rk}. ML technique is also
applied extensively to classify the jets initiated by quarks and
gluon~\cite{Lonnblad:1990bi,PhysRevD.44.2025,Andrews:2019faz,8,9,10,11,Cogan:2014oua,Lee:2019cad}.
The similar algorithm can be used for flavor tagging to distinguish the jets
initiated by heavy or light quarks or
gluon~\cite{Almeida:2015jua,Guest:2016iqz,Bols:2020bkb,Goto:2021wmw,Bielcikova:2020mxw}.
The jet images are also used by ML algorithm to tag electroweak~(EW) or quantum
chromodynamic~(QCD) jets~\cite{deOliveira:2015xxd,Barnard:2016qma}. The similar
technique can be used to distinguish between the electroweak
bosons~($W^+/W^-$,$W/Z$)~\cite{Chen:2019uar}, this can be further utilised for
studying the full hadronic channel of $W^+W^-$ which receives a large background
contribution from $ZZ \to $ hadrons. While studying BSM models or finding rare
particles which requires solving difficult signal-versus-background
classification problems, ML techniques are
used~\cite{Baldi:2014kfa,Chakraborty:2019imr,Datta:2019ndh,CMS:2019dqq,Alimena:2020web,Bernreuther:2020vhm,Cogollo:2020afo,Grossi:2020orx,Ngairangbam:2020ksz,Englert:2020ntw,Freitas:2020ttd,Freitas:2019hbk}.
We can also use such ML methods to correctly identify the nature and properties
of outgoing particles from a high energy
collisions~\cite{deOliveira:2018hva,Belayneh:2019vyx,Qasim:2019otl}. In our
study, the $W^+$ boson decays to light quarks~(\text{u$\bar{d}/c\bar{s}$}) and
due to the similar signature of the jets produced from this light quarks it is
non-trivial to tag the jets efficiently. The energy profile do acts as a good
classifier providing approximately $75\%$ accuracy in classification. Since,
energy of a final state fermions depends on the polarization of $W$ boson, we
cannot use energy as a variable to tag the jets. We tried to classify the jets
on event-by-event basis into two class based on the processes used: ($i$) $c$
vs. $\bar s$ for $W^+\to c\bar s$ sample, ($ii$) $u$ vs. $\bar d$ for 
$W^+\to u\bar d$ sample, ($iii$) $cs$ vs. $ud$, and ($iv$) $cu$ vs $sd$ for
combined sample. We train separate ANN and BDT models for each of the four cases.

We used {\tt MadGraph5$\_$aMC$@$NLO}~\cite{Alwall:2011uj} for event generation. The process generated is
			\begin{equation}
				e^- + e^+ \to W^- + W^+, W^- \to l^- \bar{\nu_l}, W^+ \to jj
			\end{equation}
			where $j$ are the light quarks~($u,d,c,s$) at an center of mass
energy $\sqrt{s}=250$ GeV. After the matrix elements events are generated, those
events are passed to {\tt Pythia8}~\cite{12} for showering and hadronization.
The final state particle is selected with $p_T\geq 0.3$ GeV and with $|\eta| \le
3.0$. The lepton from the decay of $W^-$ are excluded from further analysis. We
have used two set of jet clustering algorithm in {\tt FastJet}~\cite{13,14},
firstly the final set of particles are clustered using {\tt anti-$k_T$} jet
clustering~\cite{15} with jet radius $R=0.7$ and those {\tt anti-$k_T$} jets are
further clustered using {\tt $k_T$} algorithm~\cite{Catani:1993hr} with jet
radius $R=1.0$, such that the excluded soft particles are also clustered. It is
seen that these combination of clustering reduces the number of jets in each
events. We have found that the first two hardest jets amounts to approximately
$90\%$ of all jet momentum in each event, so it makes perfect sense to work 
with just two hardest jets rather than the whole sets of jets. To put {\em 
truth} label on the jet for supervised learning, we used
the geometric distance $\Delta R_{ij}$ between reconstructed jets and the
initial partons~(quarks). Considering only two hard jets, there are four
combination 
\begin{itemize}
				\item hardest jet near up-type quark.
				\item hardest jet near down-type quark.
				\item both jet near up-type quark.
				\item both jets near down-type quark
			\end{itemize}
\begin{figure}[h!]
\subfigure{\includegraphics[width=0.235\textwidth]{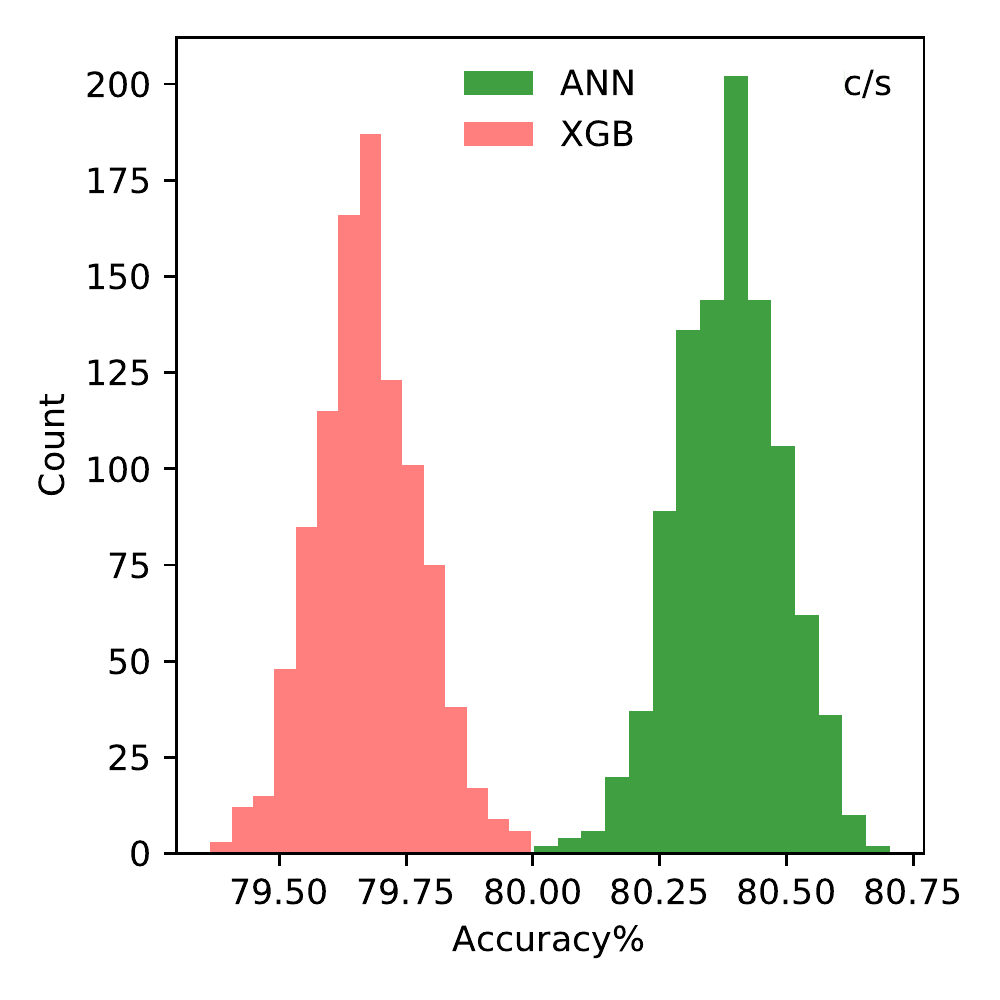}}
\subfigure{\includegraphics[width=0.235\textwidth]{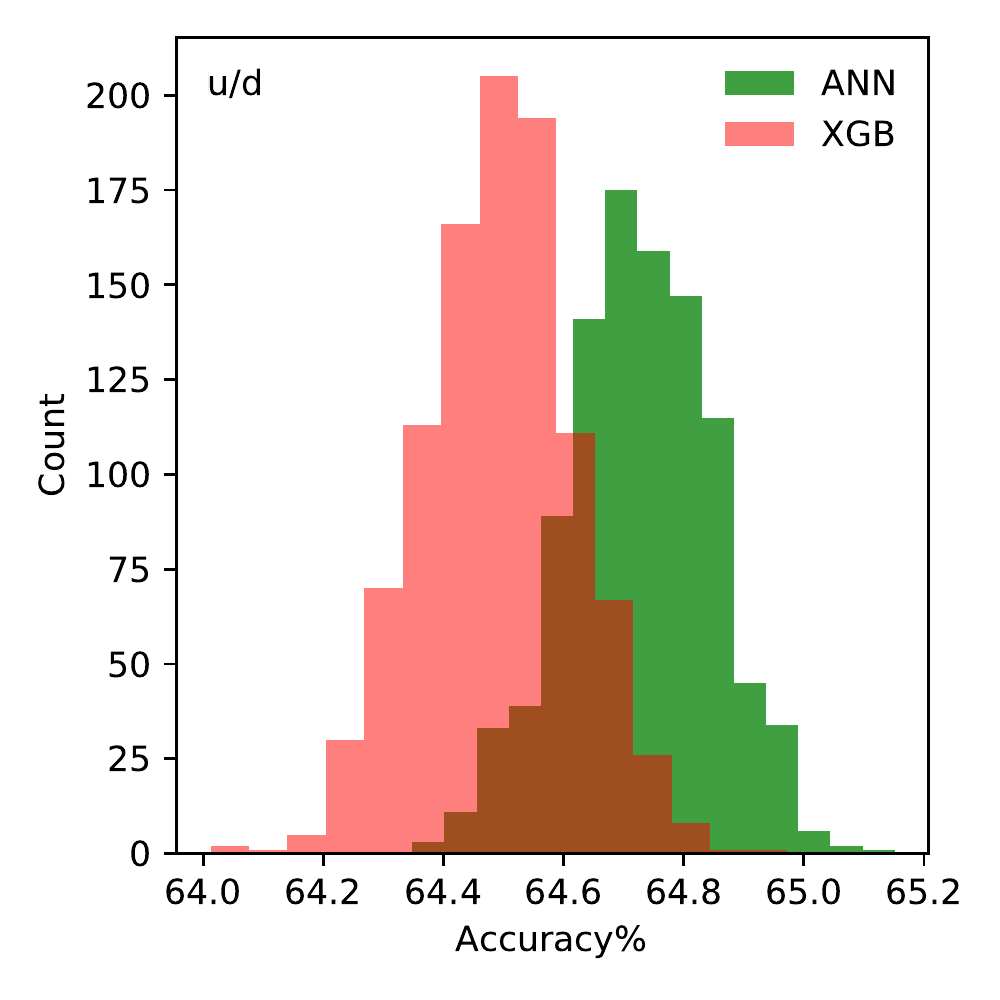}}
\subfigure{\includegraphics[width=0.235\textwidth]{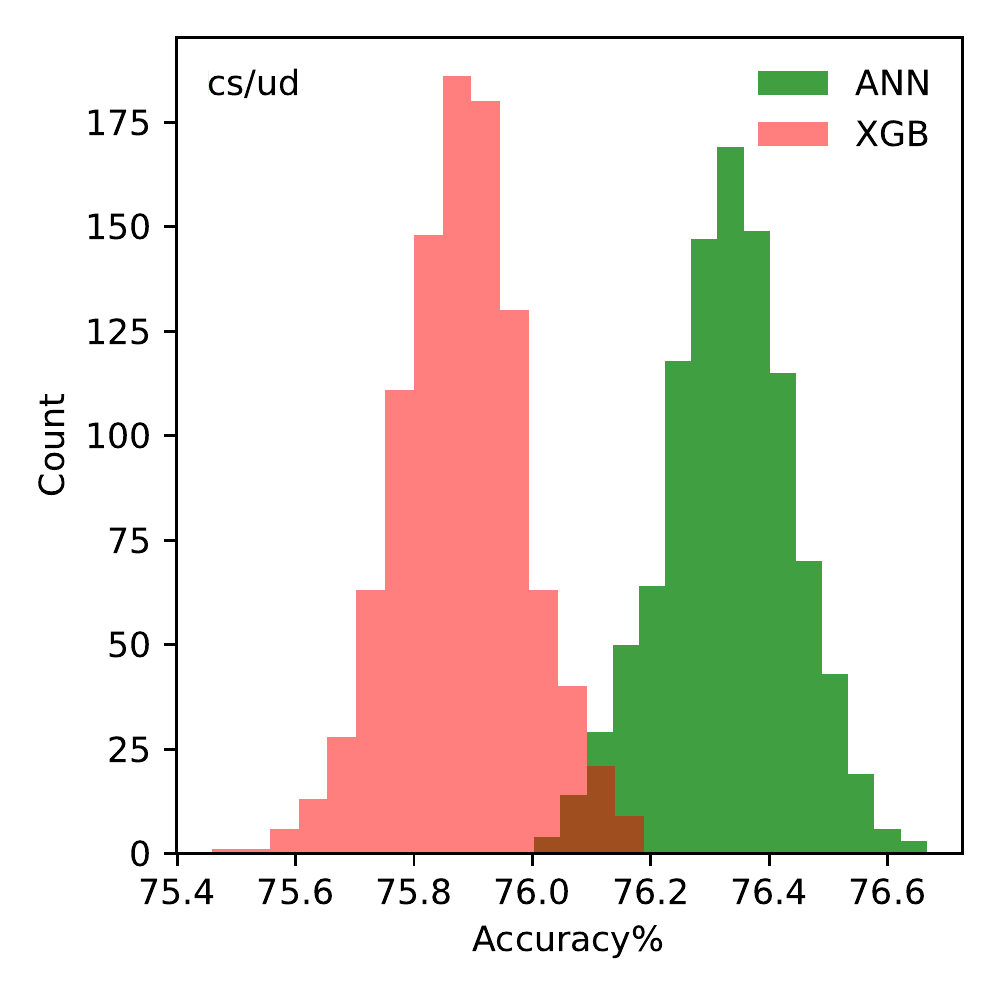}}
\subfigure{\includegraphics[width=0.235\textwidth]{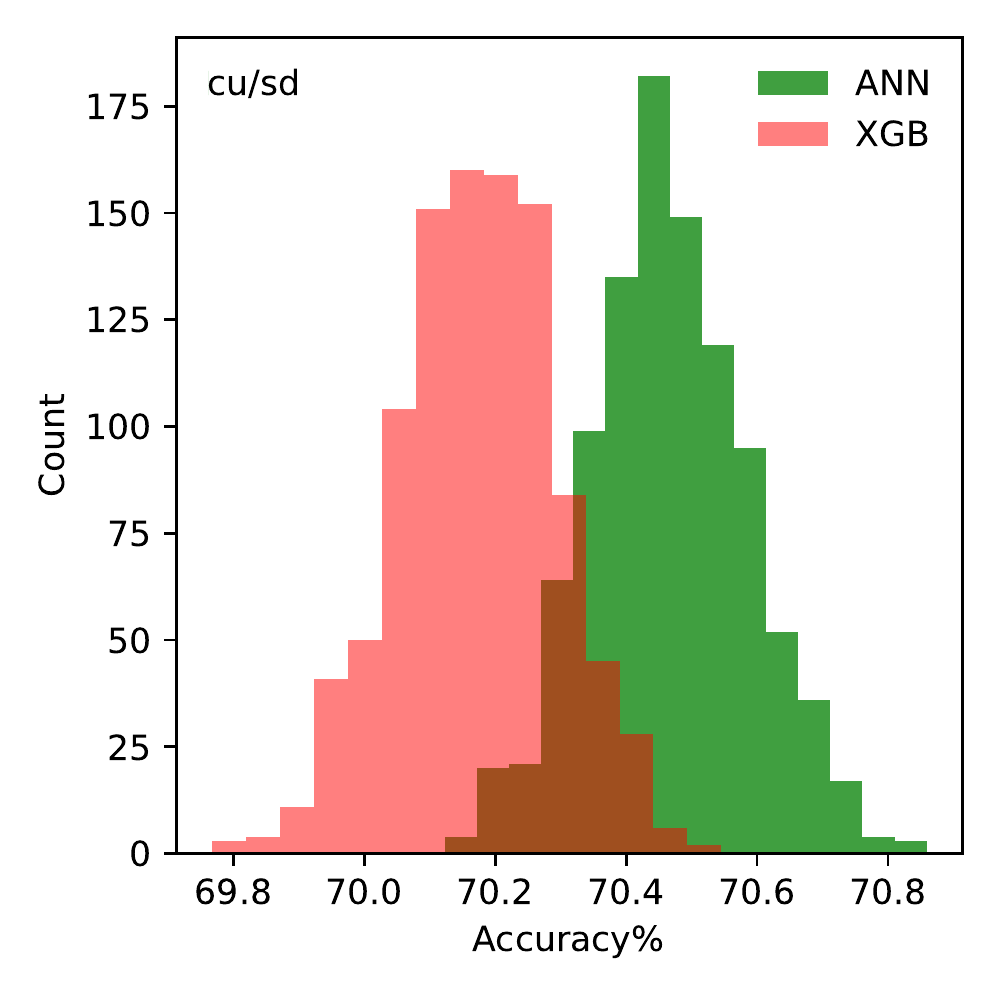}}
\caption{\label{acc:ML}Accuracy obtained using ANN and XGB for different combination of jet.}
\end{figure}
The first two condition is straight-forward, but for any event satisfying third
or fourth condition, we always set the hardest jet to the respective quark. Once
the truth labelling is set, we obtained various parameters of jets and the 
particle within the jets to make them as an input for different ML models. 
We choose following features for each of the jets:
\begin{itemize}
				\item Total number of leptons~({\tt nlep}).
				\item Total energy of positive and negative leptons~({\tt el$+$,el$-$}).
				\item Total number of hadrons~({\tt nhad}), total energy of hadrons({\tt ehad}).
				\item Total number of charged hadrons~({\tt nChad}) and the total energy of positive and negative hadrons~({\tt ehad$+$,\tt ehad$-$}).
				\item Total number of charged particles~({\tt nch}).
				\item Total number of positive and negative charged particles~({\tt nch$+$,\tt nch$-$}) and their total energy~({\tt ech$+$,\tt ech$-$}).
				\item Total number of visible particles~({\tt nvis}) and their total energy~({\tt evis}).
				\item Total energy of the photons~({\tt egamma}).
				\item \textit{Displaced Tracks:} Particles like $\Lambda^0$ baryon~($\tau = 2.631 \times 10^{-10}$ s), $K_s^0~(\tau = 8.954 \times 10^{-11}$ s)~\cite{Workman:2022ynf} have a significant lifetime, when decaying to a pair of oppositely charged daughters in case of neutral unstable mother or a charged and neutral daughter in case of charged unstable mother would produce a measurable secondary displaced tracks. The $\Lambda_c$ and $\Lambda_b^0$ would also give rise to a displaced tracks. The tracks are binned with their decay length~($\lambda = c\beta\gamma\tau$) mm:\begin{itemize}
					\item {\tt c1}: $\lambda \in \left[0.3,3.0\right]$
					\item {\tt c2}: $\lambda \in \left[3.0,30.0\right]$
					\item {\tt c3}: $\lambda \in \left[30.0,300.0\right]$
					\item {\tt c4}: $\lambda \in \left[300,1200.0\right]$
					\item {\tt c5}: $\lambda \in \left[0.3,1200.0\right]$
					\item {\tt c6}: $\lambda \in \left[1200.0,\infty\right]$
				\end{itemize} 
				\item Total number of charged tracks with non-zero impact parmaeter~({\tt tip}).
				\item A pair of charged tracks can meet at a point indicating a common mother. For such cases, we also count the number of tracks with positive and negative impact parameter separately.
				\item Total count of secondary displaced vertex~({\tt sdv}) are constructed by noting a pair of charged particles coming from a displaced point from primary vertex.
				\item The transverse mass~({\tt mTj}) of the jet.
			\end{itemize}
			The above set of discrete and continous variables are used to train our ML models. The Pearson's correlation coefficients of the variables with the jet labels~($cu \equiv 1, \bar{s}\bar{d} \equiv 0$) is shown in Table~\ref{ml_corr}.  
			\begin{table}[h!]
				\centering
				\caption{Table showing the Pearson's correlation of features with the label in $cu/\bar{s}\bar{d}$ case. The table shows only those features which has a correlation of $5\%$ and higher.}
				\label{ml_corr}
				\begin{tabular}{|p{1.5cm}|p{2.0cm}|p{1.5cm}|p{2.0cm}|}
					\hline
					Features & CorrCoeff. & Features & CorrCoeff.  \\ \hline
					{\tt mTj} &-27.0&{\tt c2}&13.7 \\ \hline 
					{\tt c6}&9.7&{\tt nvis}&-10.0 \\ \hline
					{\tt nch$+$}&-5.0&{\tt nch$-$}&6.0 \\ \hline
					{\tt nch}&10.0&{\tt nlep}&11.4 \\ \hline
					{\tt nChad}&-10.0&{\tt nhad}&-11.0 \\ \hline
					{\tt el$+$}&16.0&{\tt ehad}&-9.0 \\ \hline
				\end{tabular}
			\end{table}
			The table lists only those variables which has more than $5\%$ correlation with the jet class. For training ANN and BDT models, we generate 10 million events and separate set of one million events are used for testing the models. We used {\tt Keras} with {\tt TensorFlow} as backend to implement ANN while we used {\tt XGBoost}~\cite{16} to implement BDT. The parameters for BDT are chosen as follows:
		\begin{itemize}
			\item Learning rate, $\eta = 0.01$
			\item Maximum depth = 6
			\item L2 Regularization = 1
			\item Number of gradient boosted trees = 100
		\end{itemize} 
			The architecture of ANN is given in Table~\ref{ANN:arch}. The optimization of ANN model is done using {\tt Adam}.\\
			\begin{table}[h!]
				\caption{\label{ANN:arch}
					The table shows the architecture of artificial neural network used. It contains two hidden layer~(Layer 1 and Layer 2) and in each layer different number of nodes is used and is given in column 2. Activation function used for eachlayer is given in third column. We used weights as Glorot-Normal in our first hidden layer.}
				\begin{tabular}{|c| c |c |c|}
					\hline
					Layers& Nodes &Activation Function& Weights \\ \hline
					Layer 1 & 80 & Tanh & Glorot Normal \\ \hline
					Layer 2 & 40 & Tanh & \\ \hline
					Output & 1 & Sigmoid &\\ \hline
				\end{tabular}
			\end{table} 
We estimate the efficiency of the ML models by running them $1000$ times on a 
random subset of size $60\%$ of our test sample. The histograms of thus obtained efficiencies
are shown in Fig.~\ref{acc:ML} for both ANN and BDT models. The mean value of
these efficiencies is taken as our estimated efficiency of the corresponding
model.
We observe
that the accuracy obtained using two different algorithm~(ANN and BDT) overlaps
for three different cases i.e. $u$~vs.~$\bar{d}$, $c\bar{s}$~vs.~$u\bar{d}$, $cu$~vs.~$\bar{s}\bar{d}$ and for $c$~vs.~$\bar{s}$ case the two
algorithm differ by $\approx 1\%$. The first two classification are useful if
one have a perfect knowledge of event, i.e. the model making third type of
classification~($c\bar{s}$~vs.~$u\bar{d}$) is ideal. The ML model performs
worst for the classification of $u$~vs.~$\bar{d}$ events as the final state signature
obtained from the hadronization of these light quarks are nearly similar. The
$c$~vs.~$\bar{s}$ event classification is fairly good with an accuracy of $\approx 80\%$.
For our purpose of polarization variable reconstruction it is sufficient to distinguish the up-type jets with down-type jets i.e $cu$~vs.~$\bar{s}\bar{d}$ case. We choose BDT algorithm for the rest of the analyses.
			
\section{Parameter Estimation}
\label{Pest}
Apart from the cross-section we have a total of 80 asymmetries as discussed in section~\ref{spinobs}.
Of these, 44 are $CP$-even and 36 are $CP$-odd
asymmetries. We note that $W^-W^+$ production process had chiral couplings,
hence the polarizations and spin-spin correlation are dependent upon the
production angle $\theta_{W^-}$. The anomalous contribution is also $\cos\theta_{W^-}$
dependent and using that can improve the sensitivity. We achieved this by
dividing the $\cos\theta_{W^-}$ into 8 bins and construct all 81 observables in
each bin. This gives us a total of 648 observables. The value of observables
in each bin are obtained for a set of couplings and then those are used for
numerical fitting to obtain semi-analytical expression of all the observables as
a function of the couplings. For cross-section which is a $CP$-even observable,
the following paramterization is used to fit the data:
			\begin{equation}
				\begin{aligned}
					\label{cpevenfit}
					\sigma(\{c_i\}) &= \sigma_0 + \sum_{i=1}^3 c^i\sigma_i + \sum_{i=1}^5 c_i^2\sigma_{ii} + \frac{1}{2}\sum_i^3\sum_{j(\neq j)=1}^3 c_ic_j\sigma_{ij} \\ &+ c_4c_5\sigma_{45}.
				\end{aligned}
			\end{equation}
			For the asymmetries, the denominator is the cross-section and the numerator $\Delta\sigma\{c_i\} = A\{c_i\}\sigma$ is parametrized separately. For the $CP$-even asymmetries the parametrization of $\Delta\sigma$ is same as in Eq.~(\ref{cpevenfit}) and for $CP$-odd asymmetries it is done using:
			\begin{equation}
				\label{cpoddfit}
				\Delta\sigma(\{c_i\}) = \sum_{i=4}^5 c_i\sigma_i + \sum_{i=1}^3 c_ic_4\sigma_{i4} + \sum_{i=1}^3 c_ic_5\sigma_{i5}.
			\end{equation}
			 Here, $c_i$ denotes the five couplings of the dim-6 operators $c_i=\{c_{WWW},c_W,c_B,c_{\widetilde{W}},c_{\widetilde{WWW}}$\}. We define $\chi^2$ distance between the SM and SM plus anomalous point as
			\begin{equation}
				\label{eqn:chi}
				\chi^2(c) = \sum_k \sum_l \left(\frac{\mathscr{O}_k^l(c) - \mathscr{O}_k^l(0)}{\delta \mathscr{O}^l_k}\right)^2
			\end{equation}
			where $k$ and $l$ corresponds to observables and bins respectively and $c$ denotes some non-zero anomalous couplings. The denominator $\delta \mathscr{O} = \sqrt{(\delta \mathscr{O}_{stat})^2 + (\delta \mathscr{O}_{sys})^2}$ is the estimated error in $\mathscr{O}$. If an observables is asymmetries $A = \frac{N^+-N^-}{N^++N^-}$, the error is given by \begin{equation}
				\delta A = \sqrt{\frac{1-A^2}{\mathcal{L}\sigma}+\epsilon_{A}^2}
			\end{equation}
			where $N^++N^- = N_T = \mathcal{L}\sigma,$ $\mathcal{L}$ being the integrated luminosity of the collider which we will just call luminosity for rest of the article unless otherwise mentioned. The error in the cross-section $\sigma$ is given by 
			\begin{equation}
				\delta \sigma = \sqrt{\frac{\sigma}{\mathcal{L}}+(\epsilon_\sigma \sigma)^2}.
			\end{equation}
			Here $\epsilon_{A}$ and $\epsilon_{\sigma}$ are the fractional systematic error in asymmetries~(A) and cross-section~($\sigma$) respectively. The analyses in our current article is done for $\sqrt{s}= 250$~GeV and luminosity,
			\begin{equation}
			\mathcal{L} \in \{100\text{ fb}^{-1},250\text{ fb}^{-1},1000\text{ fb}^{-1},3000\text{ fb}^{-1}\}
			\label{lumi}
			\end{equation}
			and the systematic errors are chosen as,
			\begin{equation}
			(\epsilon_A, \epsilon_{\sigma}) \in \{(0,0),(0.25\%,0.5\%),(1\%,2\%)\}.
			\label{syst}
			\end{equation}
			We perform analyses with each value of luminosity and for each luminosity all systematics are chosen giving us a total of 12 different analyses. 
			\subsection{One Parameter Estimation}
			In this section, the observables are obtained by varying one anomalous coupling at a time and keeping all other to zero. We describe the role of various sets of observables on obtaining the constrain on those anomalous couplings. The systematics is kept at zero for this analyses.
			It is observed that the spin-spin correlation asymmetries contributes significantly to $\chi^2$. Together with polarization asymmetries of $W^\prime s$ it enhances the overall limits.
			\begin{figure}[H]
				\centering
				\subfigure{\includegraphics[width=0.235\textwidth]{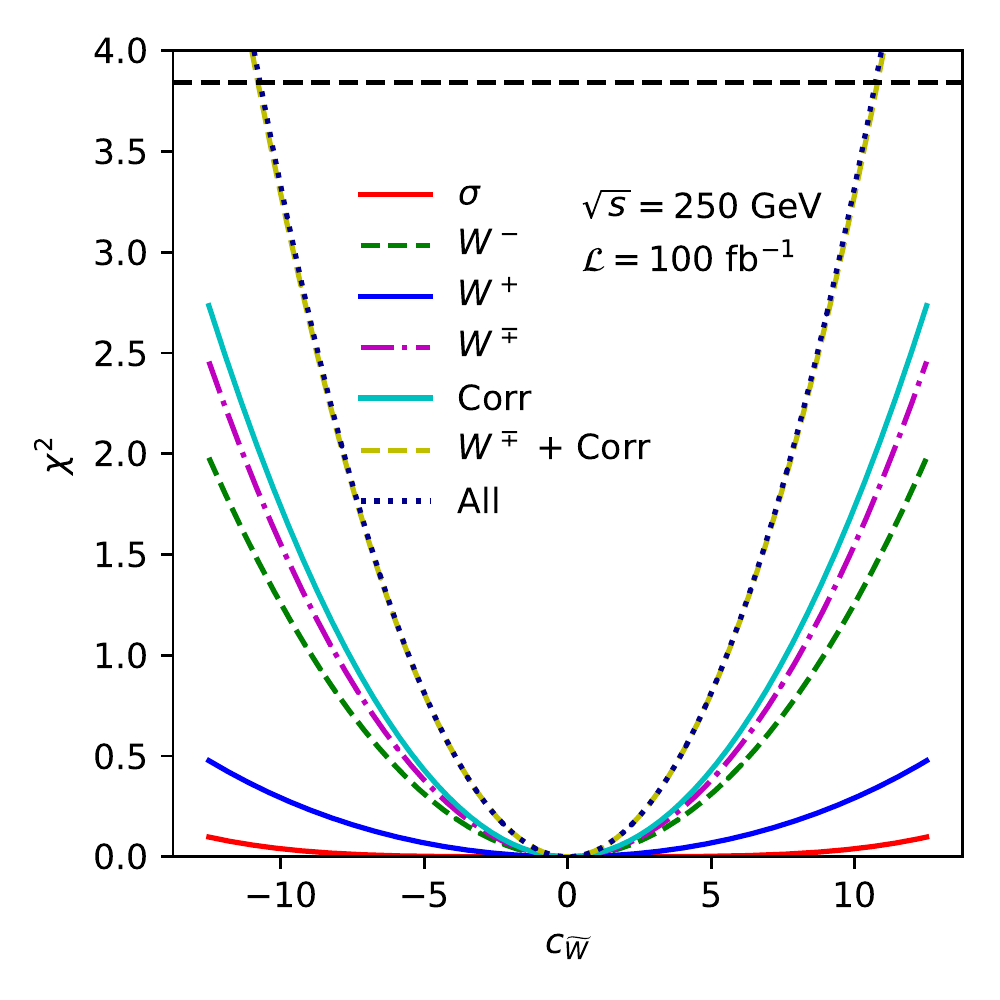}}
				\subfigure{\includegraphics[width=0.235\textwidth]{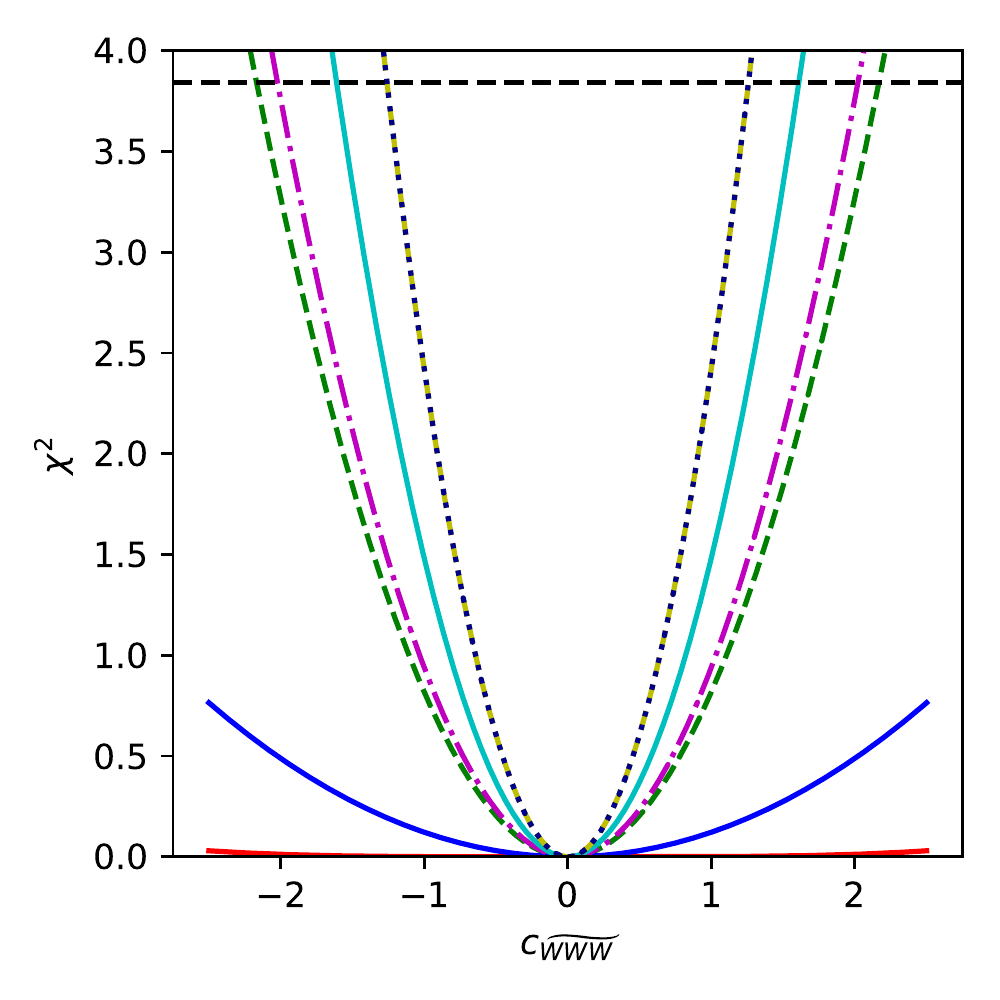}}
				\subfigure{\includegraphics[width=0.235\textwidth]{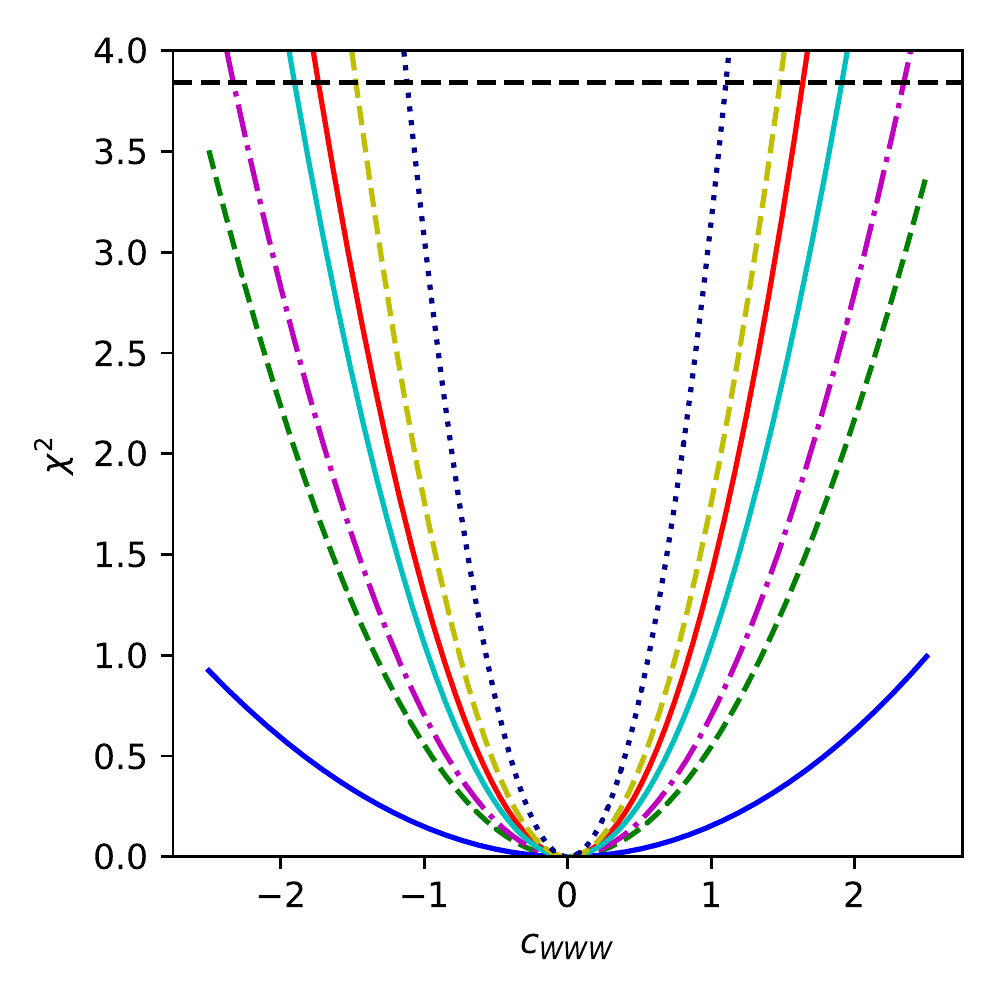}}
				\subfigure{\includegraphics[width=0.235\textwidth]{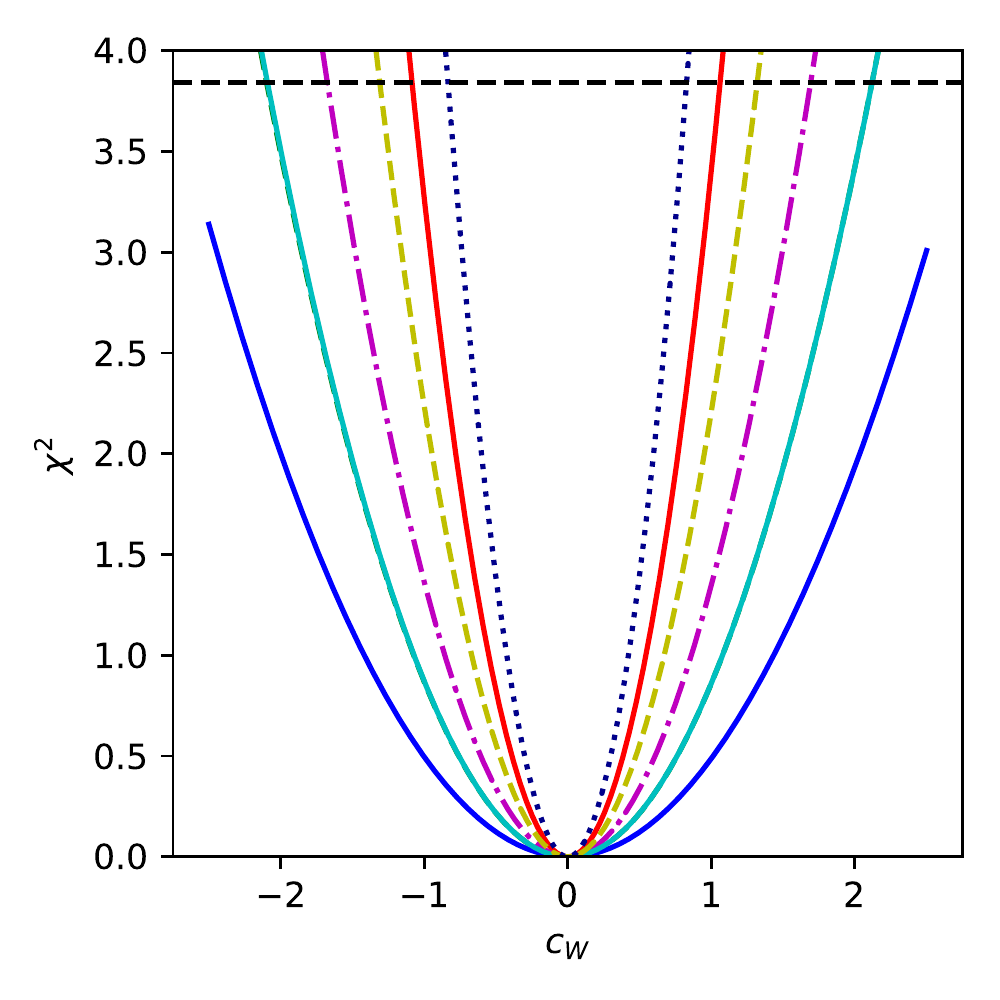}}				
			\caption{\label{1dplot}$\chi^2$  of cross-sectio~($\sigma$), asymmetries of $W$ boson and their combinations~($W^\mp$), spin-spin correlation asymmetries~(\emph{Corr}), combination of all polarization and correlation~($W^\mp+Corr$) and combination of all observables~(\emph{All}) as a function of anomalous couplings $c_i$ one at a time. The legend is kept only for $c_{\widetilde{W}}$~(right panel top row) and is same for all panel. The dashed horizontal line at $\chi^2=3.84$ indicate the 95$\%$ C.L. bound on the anomalous couplings. The systematic errors are kept to be zero.}
			\end{figure}
			In case of $CP$-even couplings like $c_{WWW}$ and $c_W$, the limits obtained by spin-spin correlation alone are approximately a factor of two tighter than the limits obtained using polarization asymmetries alone, see Fig.~\ref{1dplot} bottom row. For the $CP$-odd couplings $c_{\widetilde{W}}$ and $c_{\widetilde{WWW}}$ the limits are saturated by polarization and spin-spin correlation i.e cross-section plays little role.
			This is because the cross-section in case of $CP$-odd coupling behaves as
			\begin{equation}
				\sigma = \sigma_0 + c_{i}^2\times \sigma_i, c_i \in \{c_{\widetilde{W}},c_{\widetilde{WWW}}\}.
			\end{equation}
			And for small $c_i$, the change in cross-section is tiny. Whereas, in terms of $CP$-even couplings like $c_{WWW},c_W$ and $c_B$, the cross-section does provide a tighter limit on respective couplings because of linear dependence on $c_i$. We also note that in the case of spin-related observables, the polarization asymmetries of $W^+$ provide the smaller contribution compare to that of $W^-$; see a green curve in Fig.~\ref{1dplot} . It is because we have reconstructed asymmetries related to $W^+$ using ML models, and the reconstruction is imperfect. The best limit is obtained using all the observables together. The one parameter $95\%$ confidence level limits on various anomalous couplings $c_i$ are listed on Table~\ref{onepara95ci}.
			\begin{table}[!h]
				\centering
				\caption{\label{onepara95ci}The list of constraints on five anomalous couplings at $95\%$ confidence level obtained by varying one parameter at a time and keeping other at zero. The limits are obtained for $\sqrt{s}=250$ GeV, luminosity $\mathcal{L}=100$ fb$^{-1}$. The systematic errors are kept to be zero. }
				\begin{ruledtabular}
					\begin{tabular}{cc}
						Parameters~($c_i^{\mathscr{O}}$)&Limits~(TeV$^{-2}$) \\ \hline
						$c_{WWW}/{\Lambda^2}$&$\left[-1.12,+1.09\right]$\\
						$c_W/{\Lambda^2}$&$\left[-0.84,+0.82\right]$\\
						$c_B/{\Lambda^2}$&$\left[-2.65,+2.58\right]$\\
						$c_{\widetilde{W}}/{\Lambda^2}$&$\left[-10.76,+10.76\right]$\\
						$c_{\widetilde{WWW}}/{\Lambda^2}$&$\left[-1.24,+1.24\right]$\\
					\end{tabular}
				\end{ruledtabular}
			\end{table}\\
			The limits on anomalous couplings $c_W,c_B$ and $c_{\widetilde{W}}$ are tighter than the experimental limits quoted on Table~\ref{tab:constraint} and the limits on $c_{WWW}$ and $c_{\widetilde{WWW}}$ are comparable to the best experimental limits. 
\subsection{Two Parameter Analysis}
\begin{figure}[!h]
				\centering
				\subfigure{\includegraphics[width=0.235\textwidth,height=0.165\textheight]{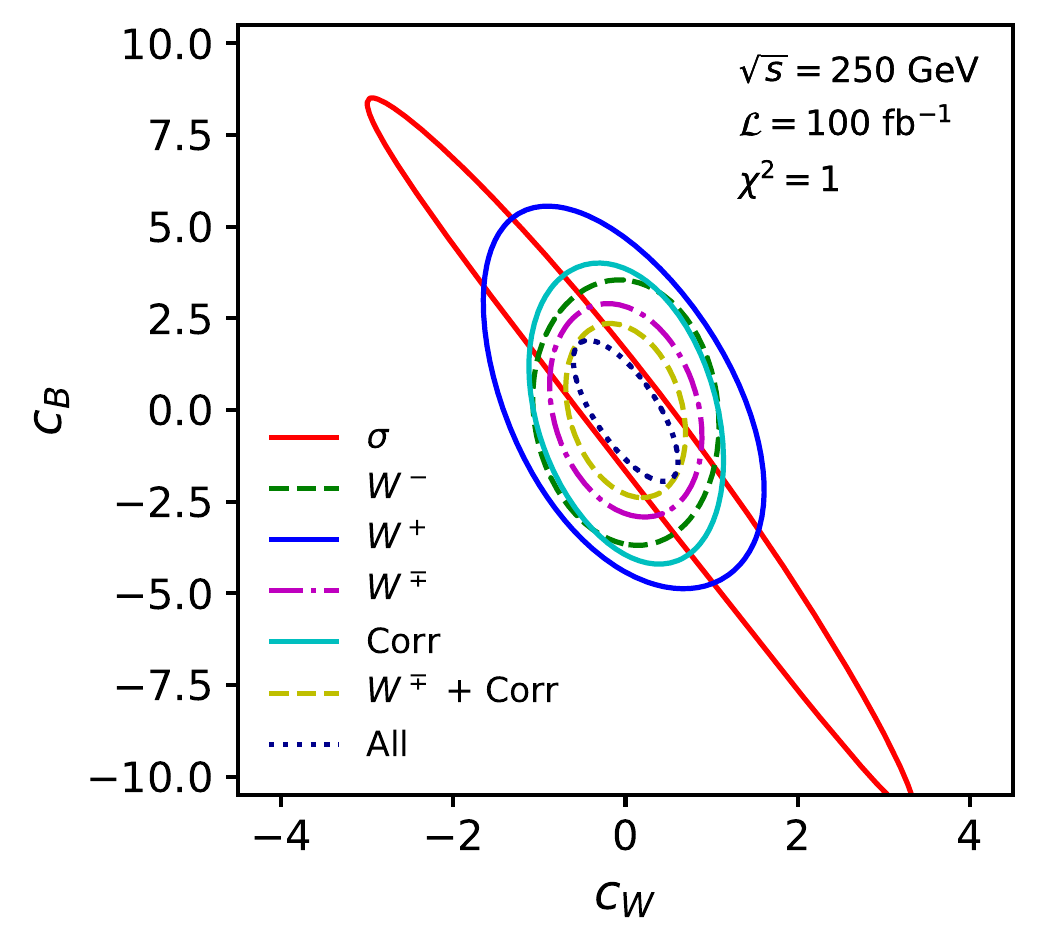}}
				\subfigure{\includegraphics[width=0.235\textwidth,height=0.165\textheight]{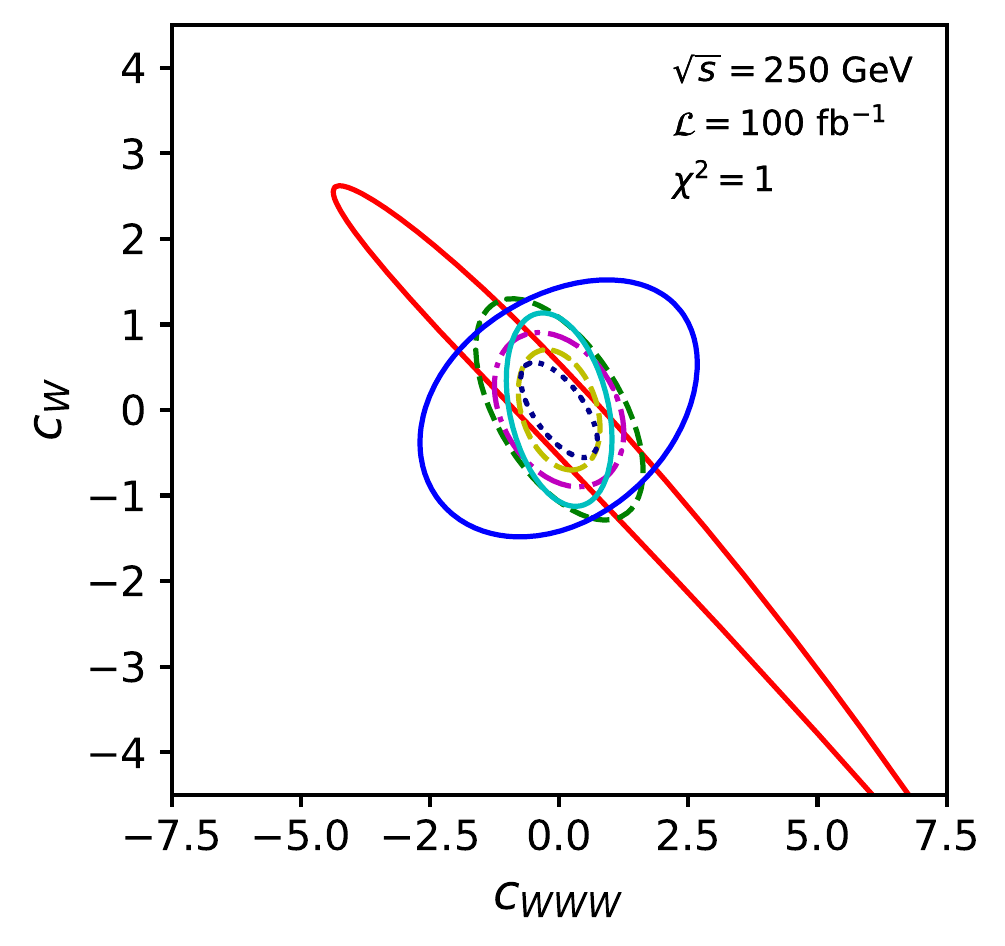}}
				\subfigure{\includegraphics[width=0.235\textwidth,height=0.165\textheight]{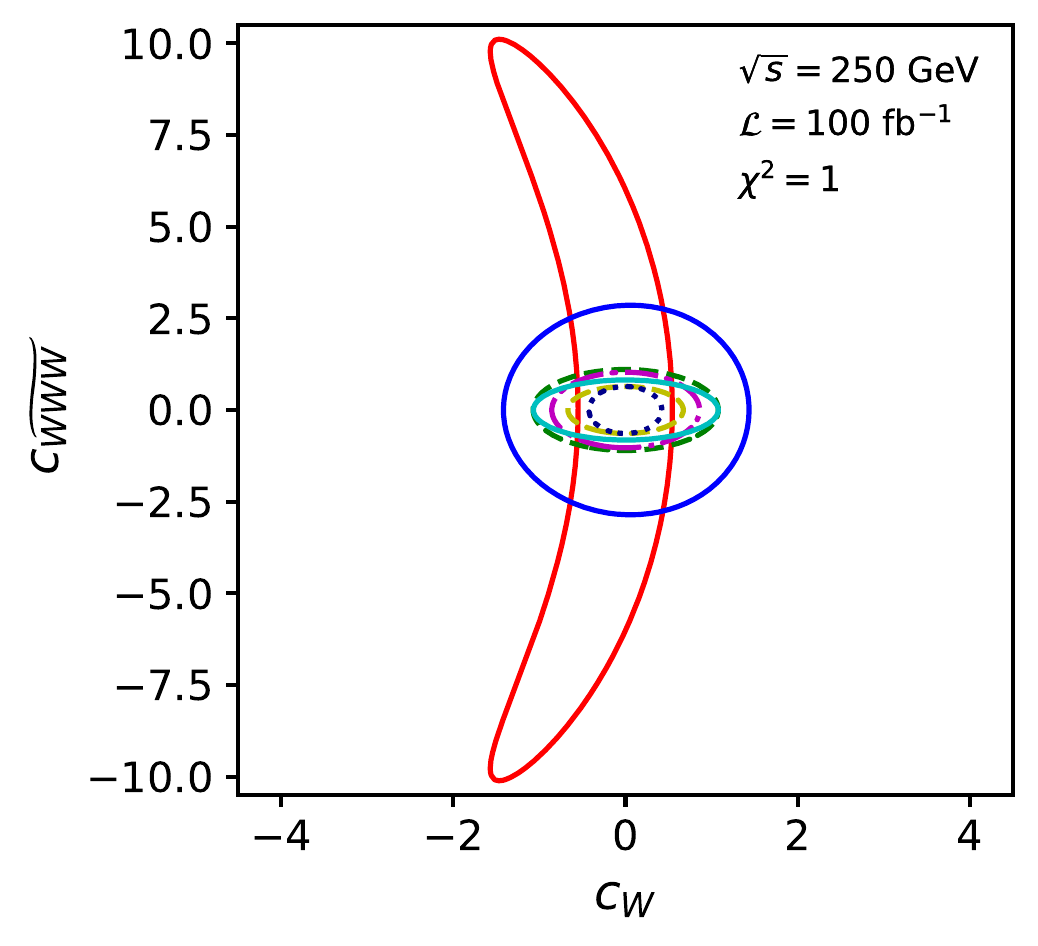}}
				\subfigure{\includegraphics[width=0.235\textwidth,height=0.165\textheight]{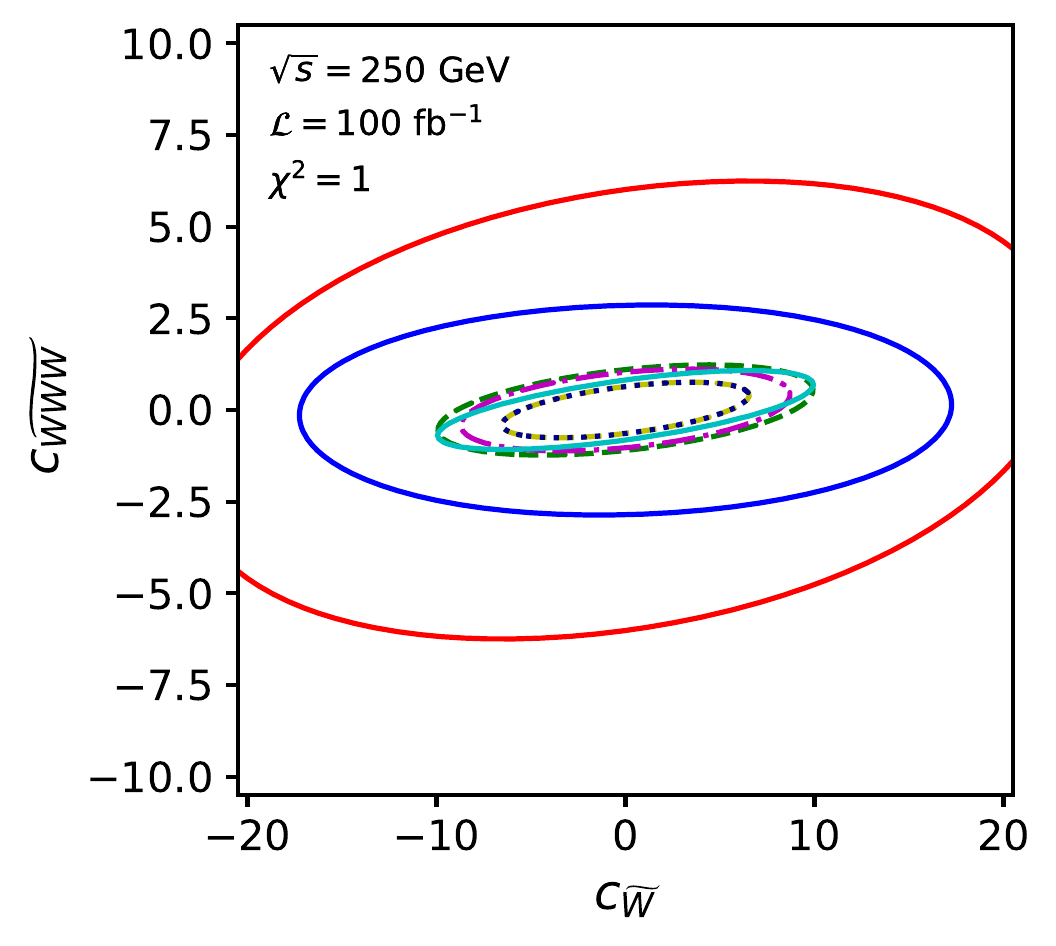}}
				\caption{\label{xgbcwwwcwcwcwt}Two dimensional contour plot showing $\chi^2 =1$ for cross-section~($\sigma$), asymmetries of $W$ boson and their combination~($W^\mp$), spin correlation~(\emph{Corr}), combination of spin related observables~($W^\mp+Corr$) and all observables~(\emph{All}) as a funciton of two anomalous couplings at a time. The legend for each panel follows that of right panel top row~($c_{W},c_B$). The systematic errors are chosen to be zero.}
			\end{figure}

			Here we discuss the case where two out of five anomalous couplings are varied at a time and others are kept at zero i.e. $(c_i,c_j)_{i\neq j}, c_i,c_j \in (c_{WWW},c_W,c_B,c_{\widetilde{W}},c_{\widetilde{WWW}})$. We study how various set of observables performs on constraining pair of anomalous couplings. Each observables corresponding to individual pair~($c_i,c_j$) is fitted to Eq.~(\ref{cpevenfit}) and Eq.~(\ref{cpoddfit}) and the fitted function are used to compute $\chi^2$ using Eq.~(\ref{eqn:chi}). We kept the systematic error at zero for this analysis. In Fig.~\ref{xgbcwwwcwcwcwt}, we show four different pairs depicting $\chi^2=1$ contour from a total of $\binom{5}{2} = 10$ pairs.
			It is observed that only certain combinations are correlated like $\{(c_W,c_B),(c_{WWW},c_W),(c_{\widetilde{W}},c_{\widetilde{WWW}})\}$ and the other remaining pairs shows little or no correlations. The contribution of various observables on constraining the anomalous couplings is shown in various panel of Fig.~\ref{xgbcwwwcwcwcwt}. The behaviour of red contours due to  cross-section $\sigma$ can be understood as follows. In terms of two parameters $\sigma$ behave as:
			\begin{equation}
				\sigma = \sigma_0 + c_1\sigma_{01}+c_2\sigma_{02} + c_{1}^2\sigma_{11}+c_2^2\sigma_{22}.
				\label{sigma2para}
			\end{equation}
			In the bottom right panel of Fig.~\ref{xgbcwwwcwcwcwt} both the parameters are $CP$-odd, i.e. the linear terms in the above equation are absent as was observed in the previous section. This leads to poor constraint on both parameters and hence the corresponding contour is large. In the top row both the parameters are $CP$-even and hence all terms of Eq.~(\ref{sigma2para}) are non-zero. The contribution from the quadratic peices, i.e. last two terms are always positive. The contribution from the linear can be of either sign depending upon the sign of the coupligns and the sign of the interference terms $\sigma_{01}$ and $\sigma_{02}$.  They lead to a vanishing contribution along the line 
			$$c_1\sigma_{01}+c_2\sigma_{02} = 0 \ \ {\rm or} \ \ c_1 = -c_2\frac{\sigma_{02}}{\sigma_{01}}.$$
			As it is clear from the figure, $\sigma_{01}$ and $\sigma_{02}$ are of the same sign and hence a poor limit along the above line in the second and fourth quadrant. The same leads to a tighter constraint in the orthogonal direction.
  			In the case of pair containing $CP$-even and $CP$-odd parameters, the cross-section as defined in Eq.~(\ref{sigma2para}) contains only one linear term coming from $CP$-even parameter, thus we see tight constrain on x-axis~($CP$-even) and loose on y-axis~($CP$-odd). The asymmetries are rational polynomials of second degree and can have complicated shapes depending upon thier $CP$ nature and the pair of couplings we are probing. The contours from the $W^+$ polarization alone are wider than the one obtained from $W^-$ as in one-parameter analyses. The contours due to correlations alone are comparable to the contribution of other observables. A combination is expected to improve the constraint. We consider all the observables for our analyses further.		    
			\begin{figure*}[!ht]
				\includegraphics[scale=0.75]{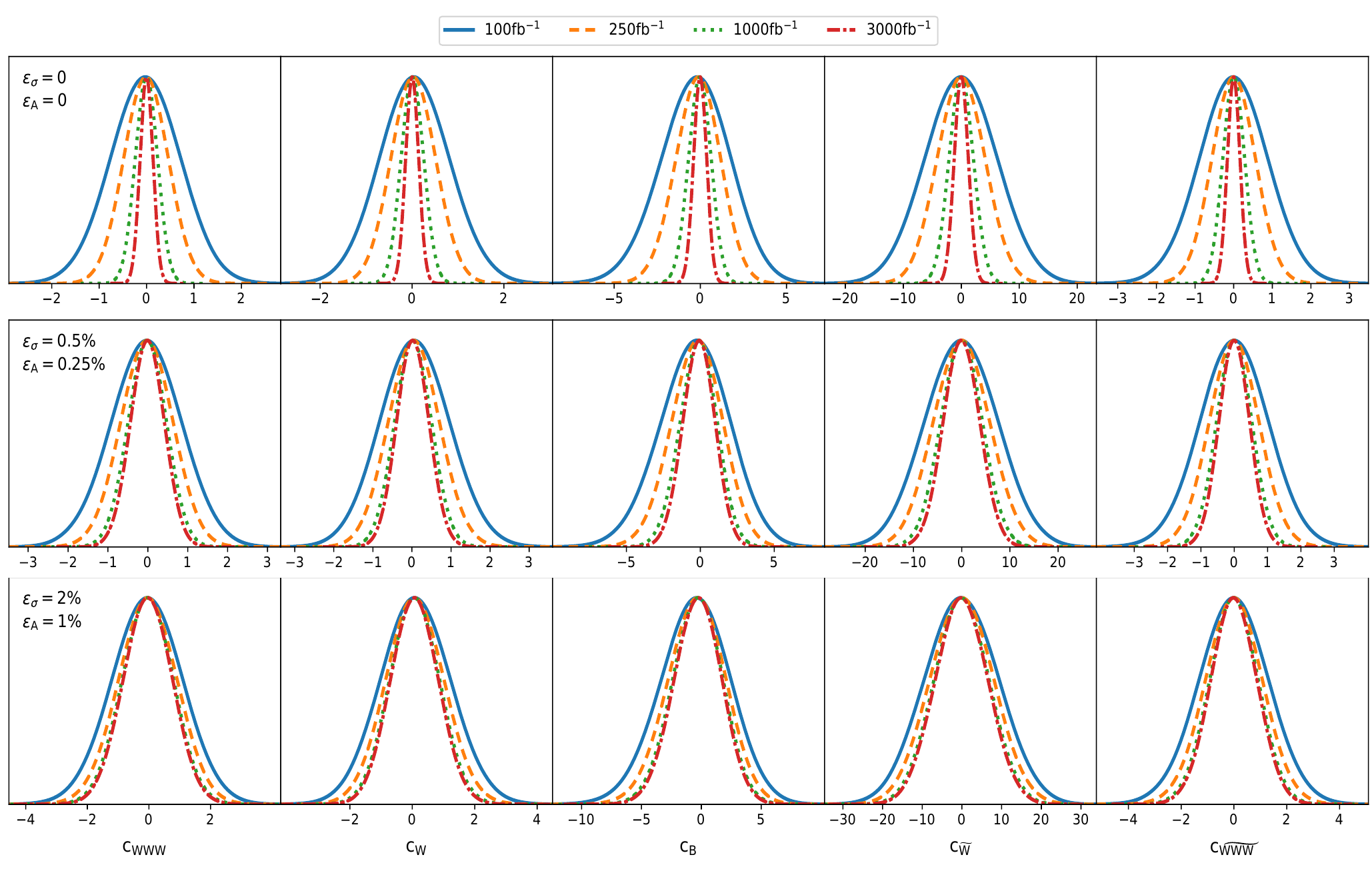}
				\caption{\label{single_95}Marginalized 1-D projections at $95\%$ C.L from the MCMC fro the effective operators(TeV$^{-2}$) for a set of systematic error and luminosities. The value of the systematics used are given in first plots of each row.}
			\end{figure*}
\subsection{Five Parameter Analysis}
\begin{figure*}[!htb]
				\includegraphics[width=0.9\textwidth]{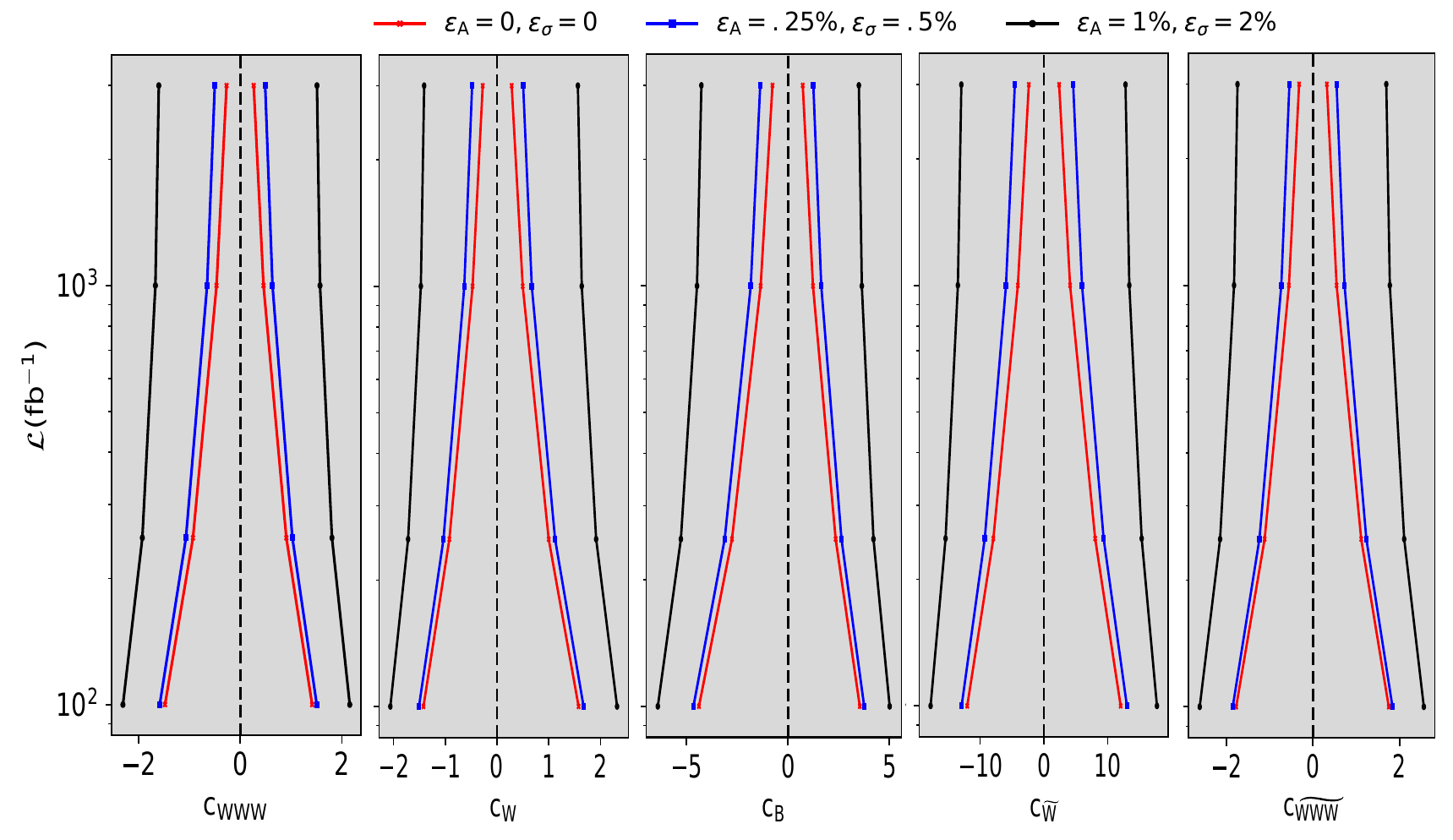}
				\caption{\label{lumvseps}The variation of the values of the five anomalous couplings~$c_i$ w.r.t luminosities $\mathcal{L}$ at different value of systematic error ($\epsilon_{A},\epsilon_{\sigma}$).}
			\end{figure*}

			Next we move to full 5-dimensional parameter space, where all parameters can vary simultaneously. We also did five parameter fitting as in Eq.~(\ref{cpevenfit}) and Eq.~(\ref{cpoddfit}) and use the fitted function to perform Monte-Carlo-markov-Chain~(MCMC)~\cite{Workman:2022ynf} analyses to estimate simultaneous limits on the anomalous couplings. MCMC is essentially Monte Carlo integration using Markov chains. These kind of integration is usually used to integrate over the high-dimensional probability distributions to make inference about model parameters or to make predictions. The chain can be constructed following the general algorithm suggested by Metropolis and Hastings~\cite{metropolis,hastings}. If $\mathcal{O}$ denotes the observed data and $\theta$ denote model parameters, we can set up joint probability distribution $P(\mathcal{O},\theta)$ over all random quantities which can be defined as 
     		\begin{equation}
				P(\mathcal{O},\theta) = P(\mathcal{O}|\theta)P(\theta)
			\end{equation}
			where $P(\theta)$ and $P(\mathcal{O}|\theta)$ are prior and likelihood distribution respectively. And Bayes theorem can be used to find the posterior distribution of $\theta$:
			\begin{equation}
				P(\theta|\mathcal{O}) = \frac{P(\theta)P(\mathcal{O}|\theta)}{\int P(\theta)P(\mathcal{O}|\theta)d\theta}
			\end{equation}
			Any features of the posterior distribution can be expressed in terms of posterior expectations of functions of $\theta$ given by 
			\begin{equation}
				E[f(\theta)|\mathcal{O}] = \frac{\int f(\theta)P(\theta)P(\mathcal{O}|\theta)d\theta}{\int P(\theta)P(\mathcal{O}|\theta)d\theta}
				\label{posterior}
			\end{equation}
			In our current analysis, we have five different parameters~($c_i$) and 648 observables as is already described.
			We defined a likelihood function by using the $\chi^2$ defined in Eq.~(\ref{eqn:chi}) and  is defined as,
			\begin{equation}
				P(c_i|\mathscr{O}) \propto e^{-\frac{\chi^2({c_i})}{2}}
			\end{equation}
			\begin{figure*}[!ht]
				\subfigure{\includegraphics[scale=0.35]{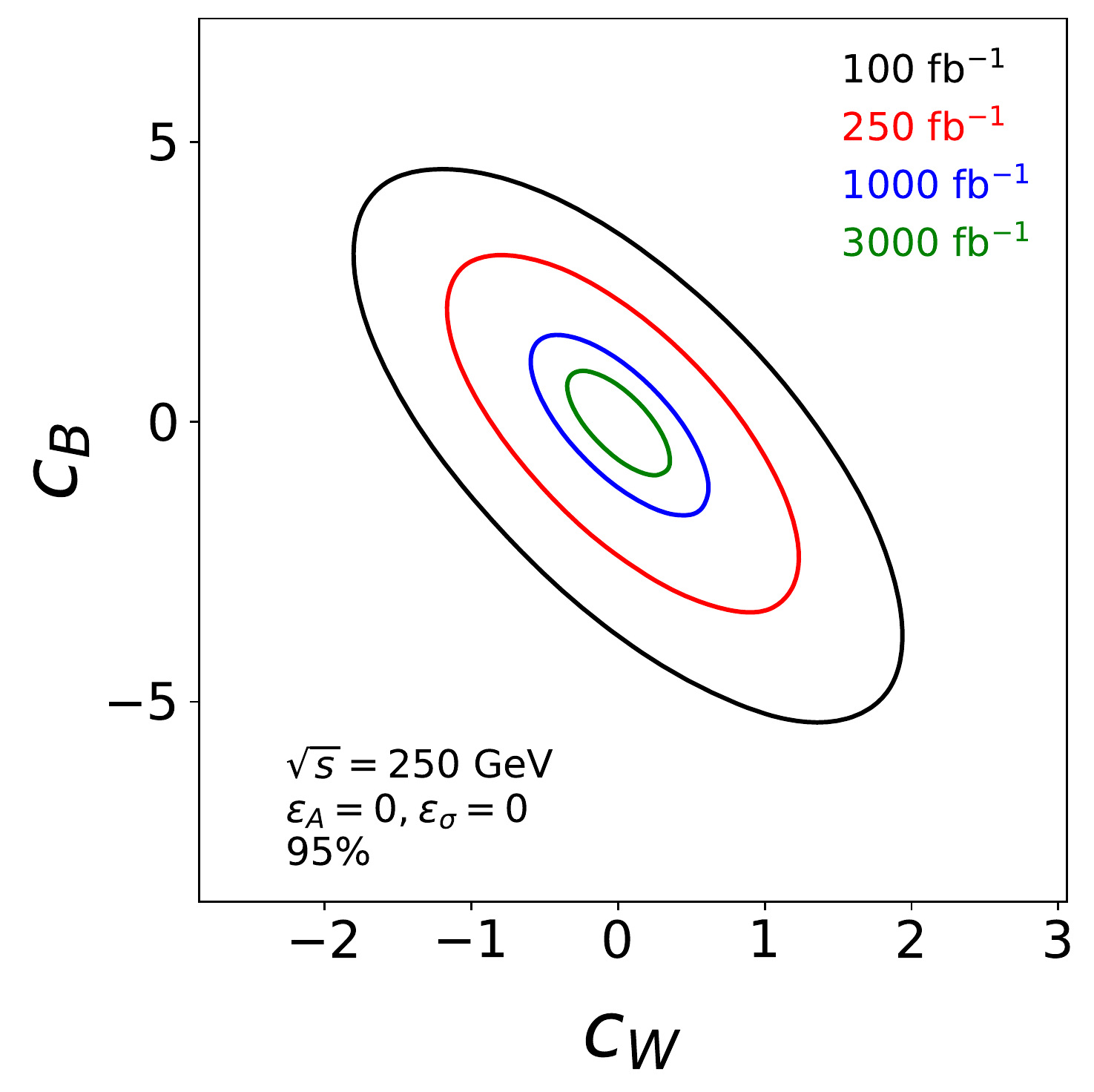}}
				\subfigure{\includegraphics[scale=0.35]{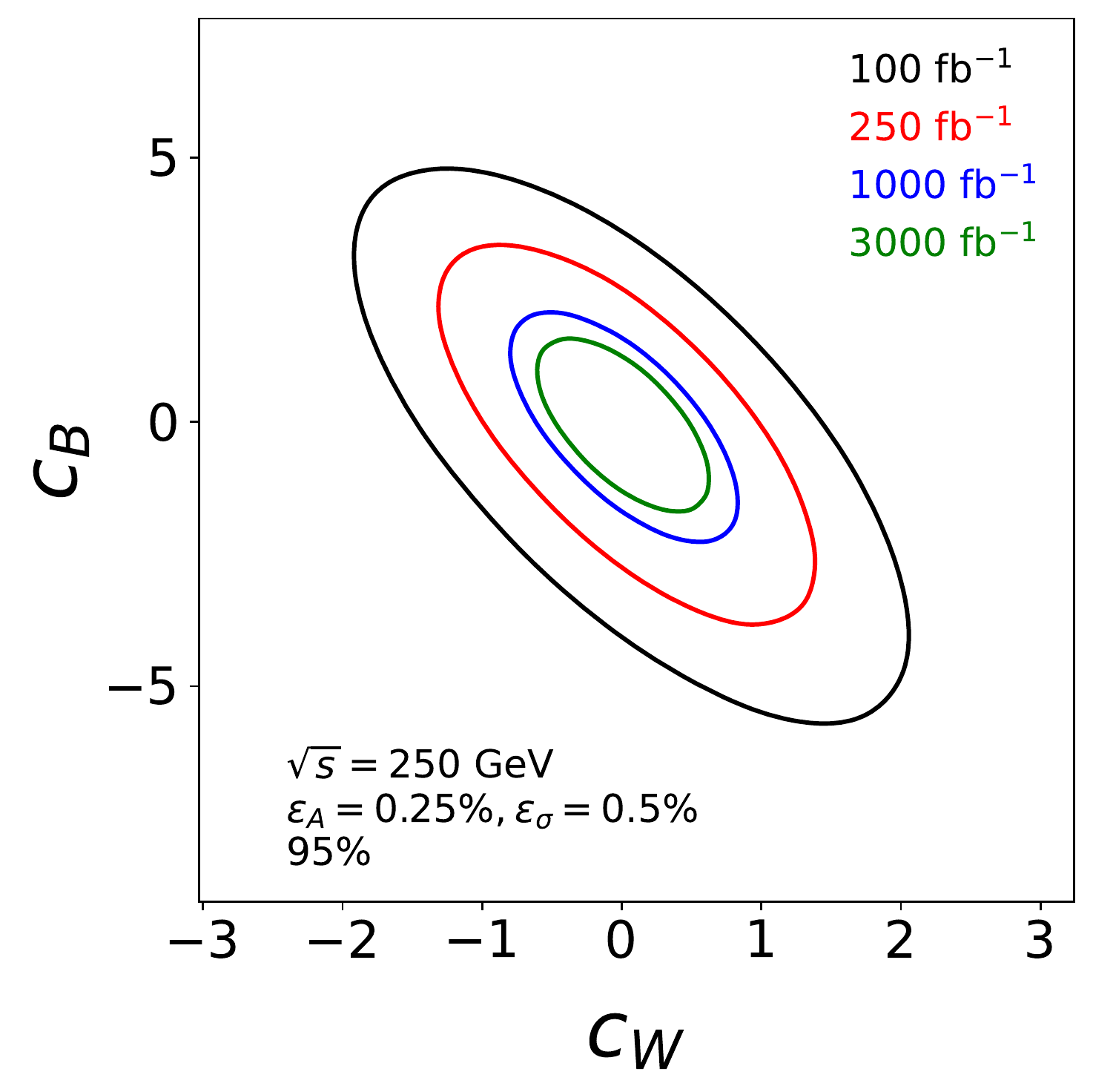}}
				\subfigure{\includegraphics[scale=0.35]{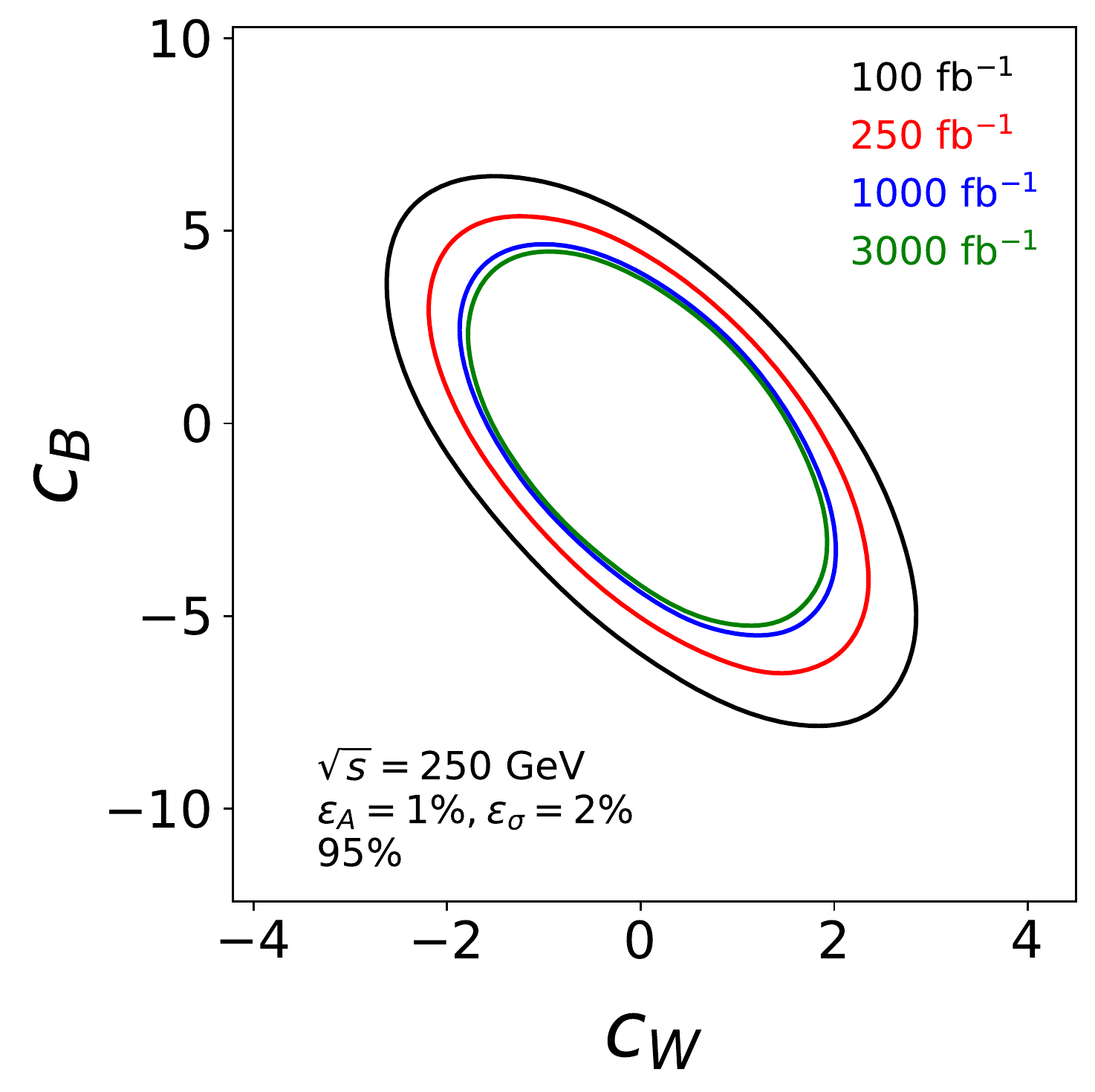}}
				\subfigure{\includegraphics[scale=0.35]{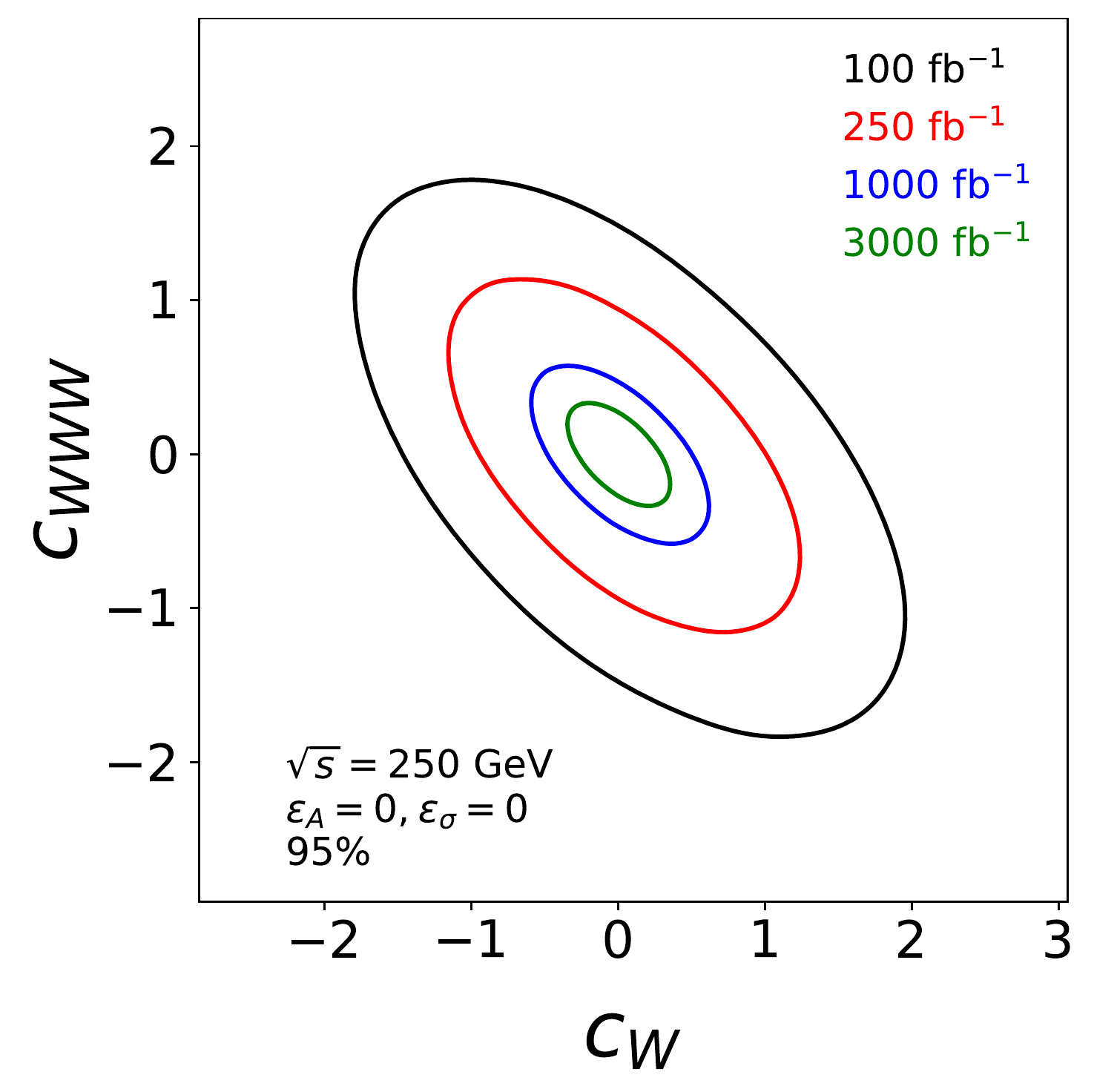}}
				\subfigure{\includegraphics[scale=0.35]{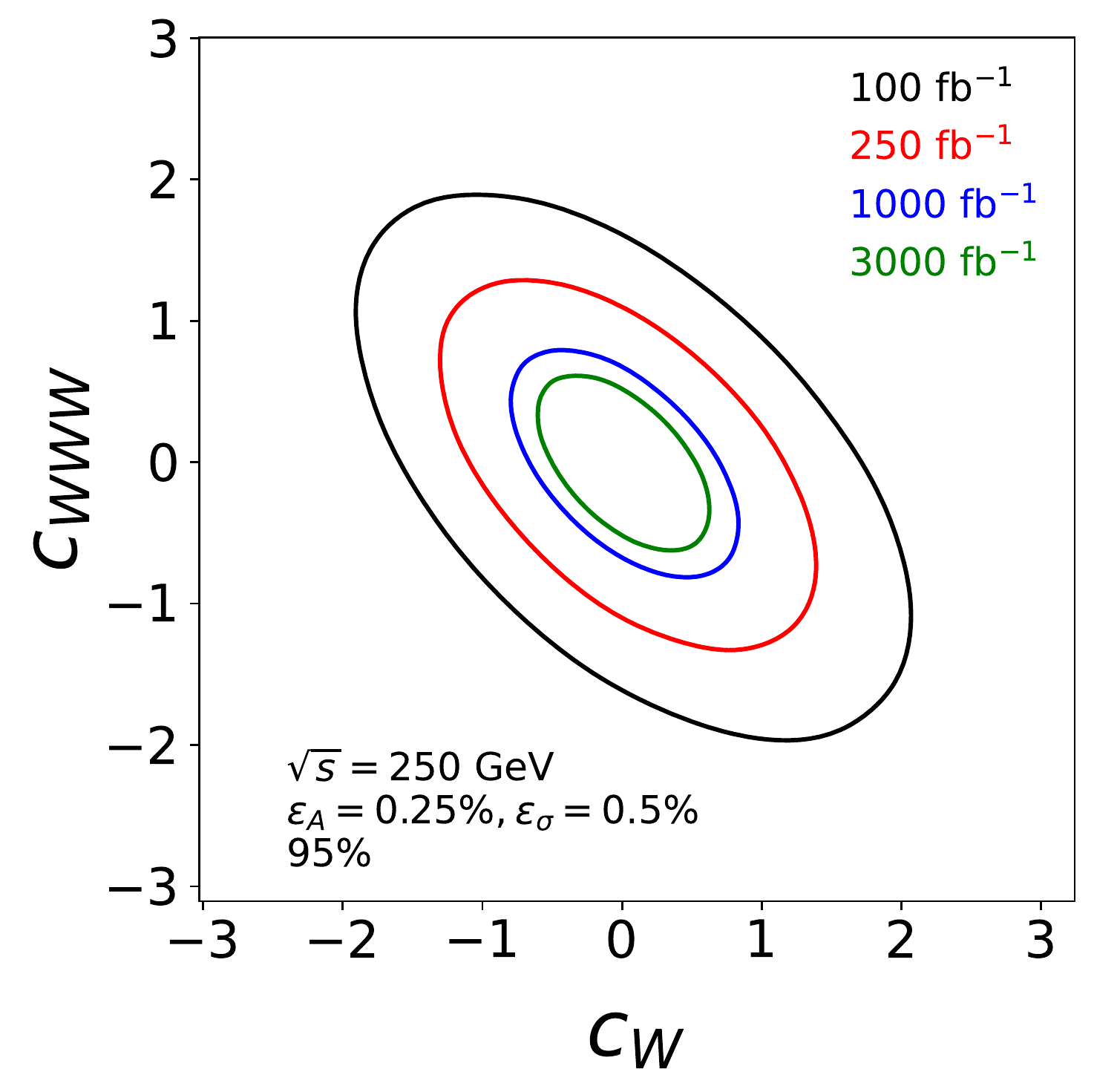}}
				\subfigure{\includegraphics[scale=0.35]{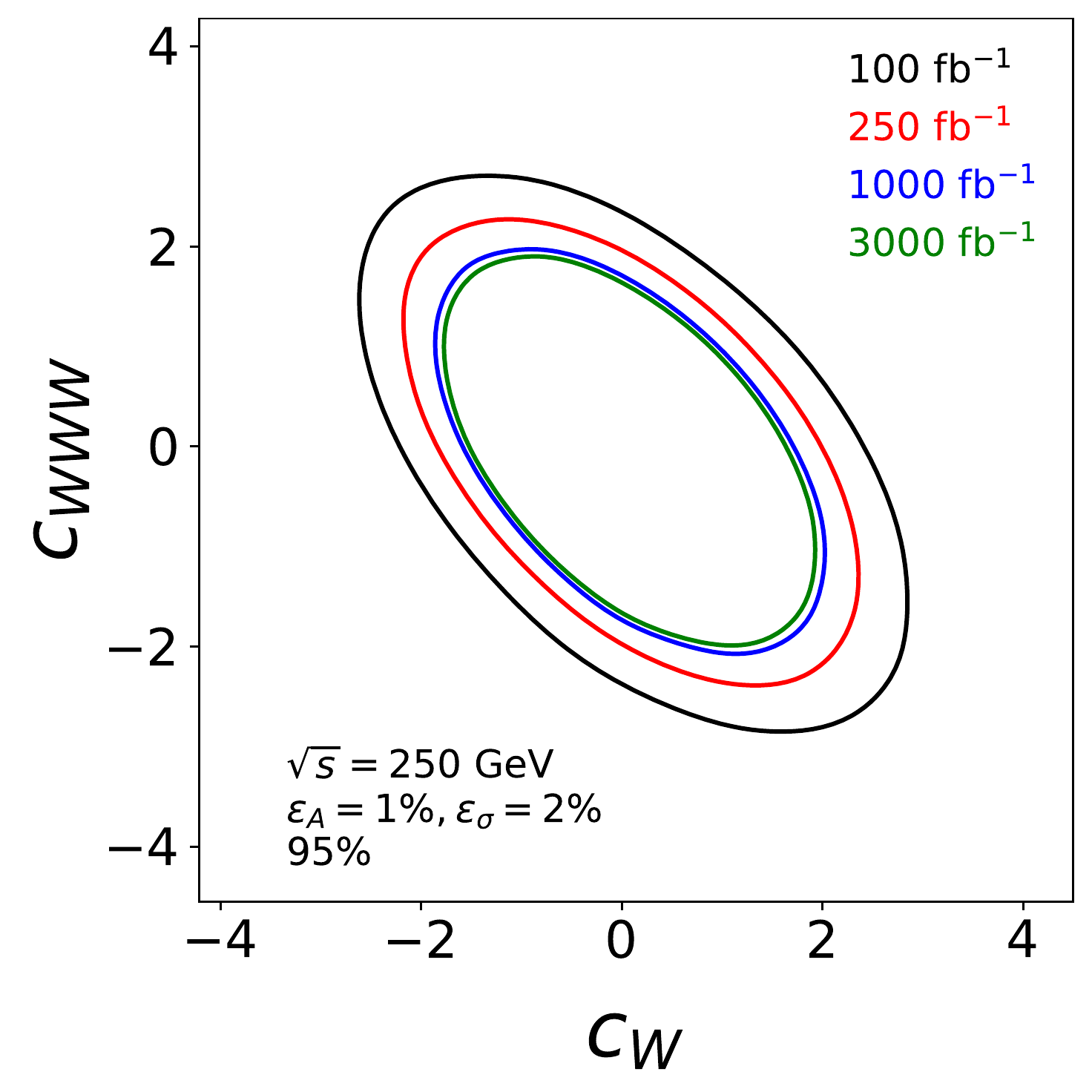}}
				\subfigure{\includegraphics[scale=0.35]{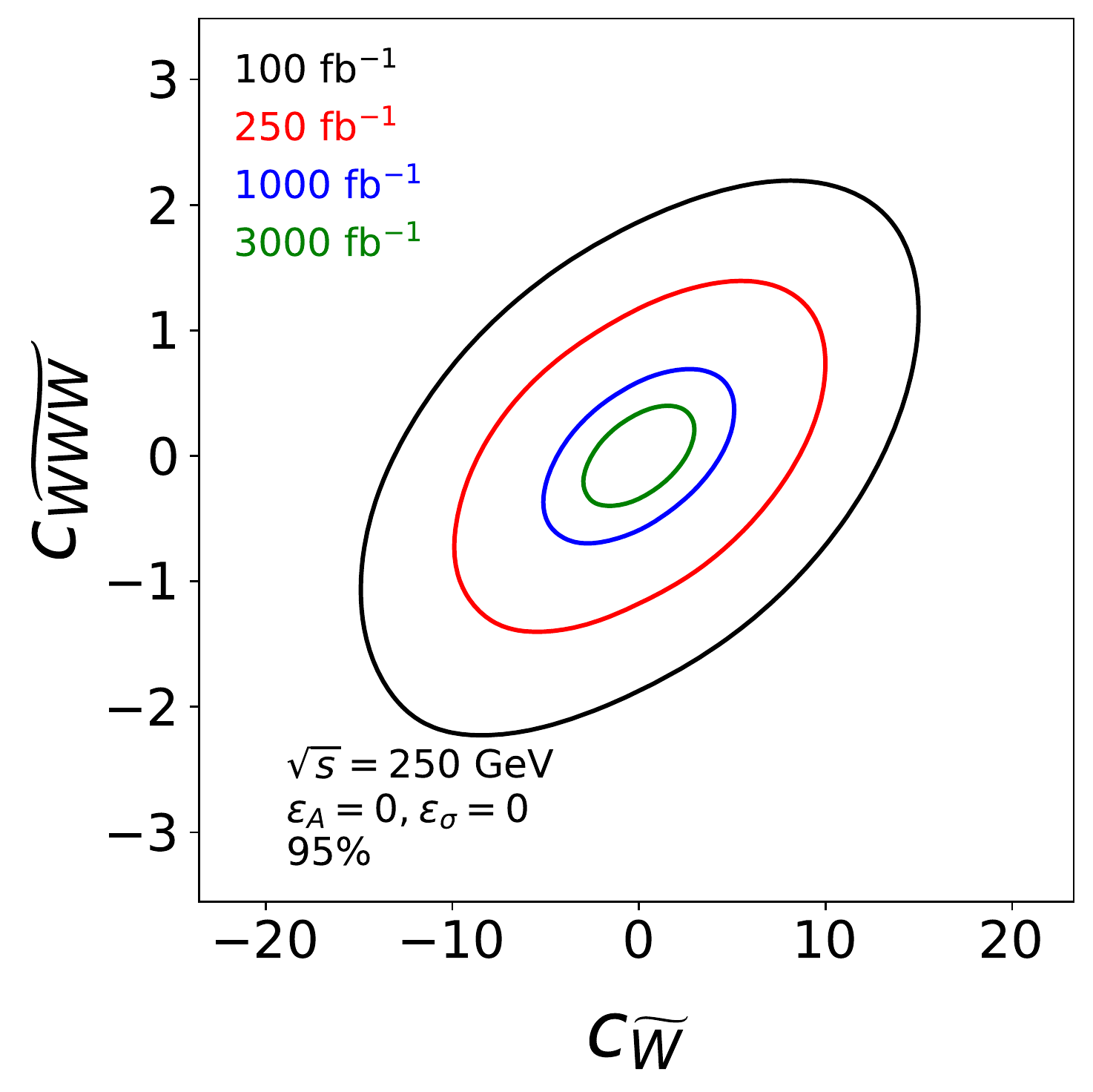}}
				\subfigure{\includegraphics[scale=0.35]{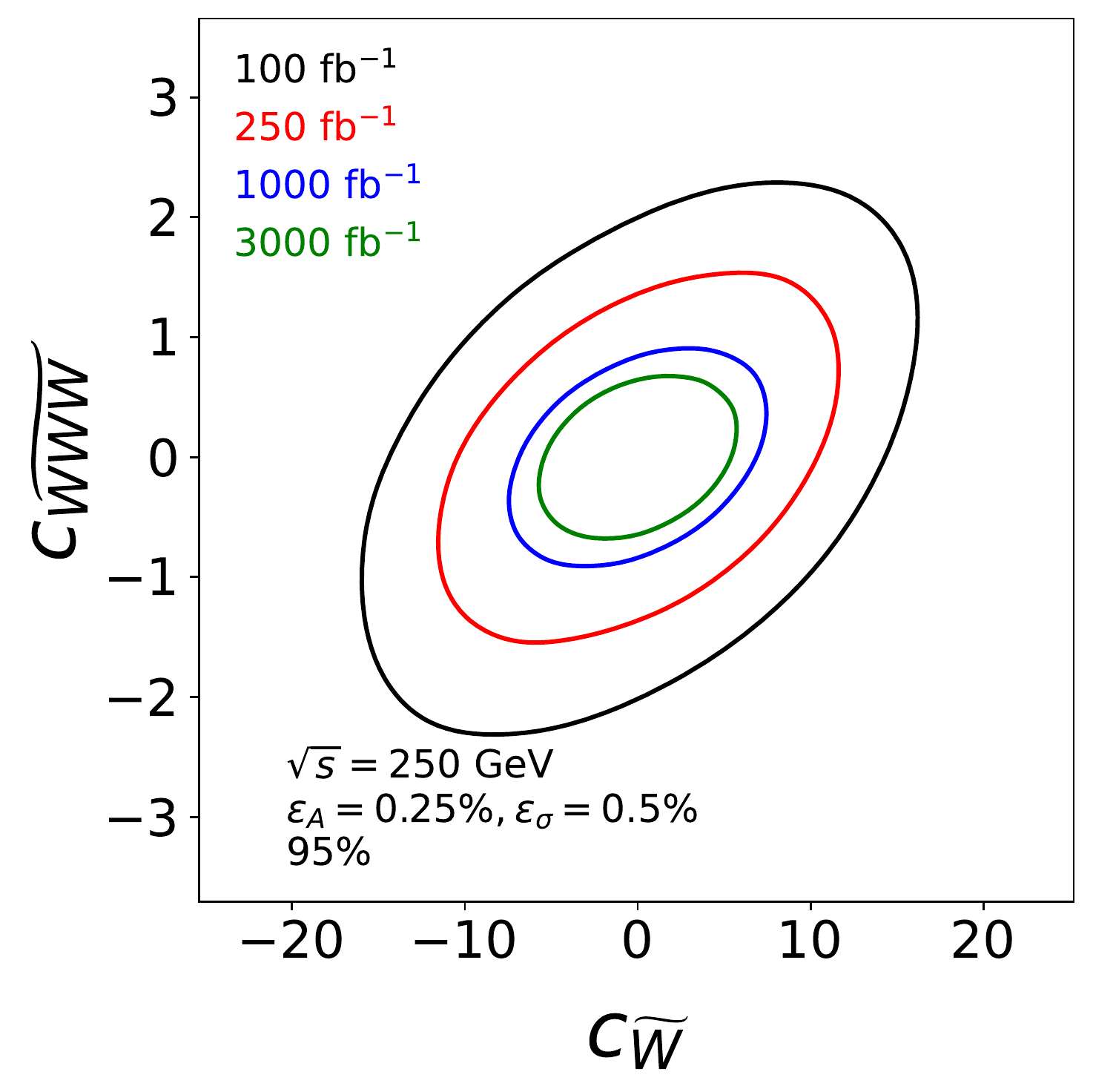}}
				\subfigure{\includegraphics[scale=0.35]{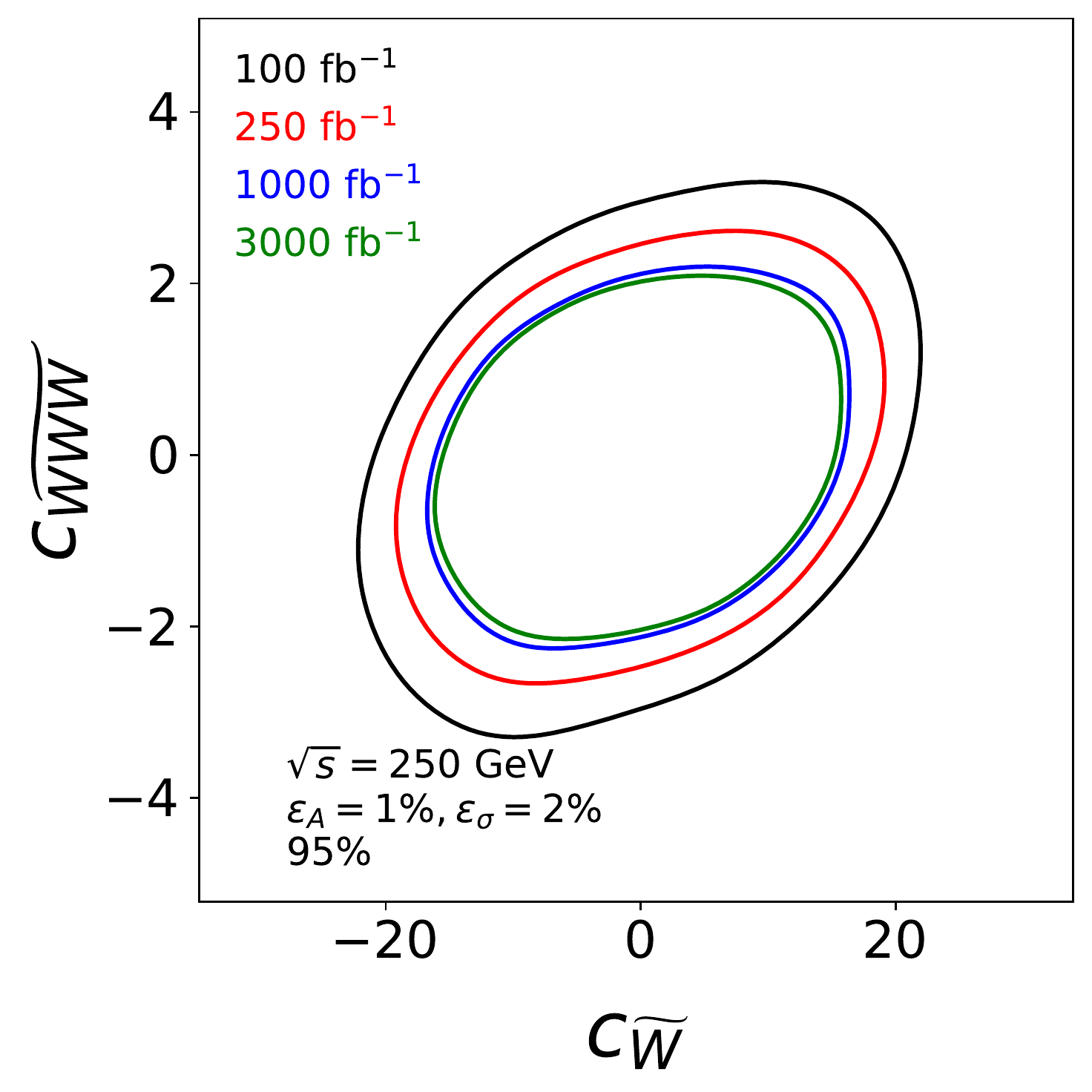}}
				\caption{Marginalized 2-D projections of the maximal correlated parameters at $95\%$ C.L. from the MCMC for a set of systematic error and luminosities obtained using  {\tt GetDist} at the  \emph{Reconstruction Level}. }
				\label{Most corr pairs}
			\end{figure*}  
			The further interpretation of the posterior probability to find the Bayesian Confidence interval is done using {\tt GetDist}~\cite{Lewis:2019xzd}. It is a python package for analysing Monte Carlo samples, including correlated samples from MCMC. The samples are obtained for a set of luminosities~$\mathcal{L}$ with different combination of systematic error of cross-section and asymmetries. For each combination of luminosity and systematic error, we generated 50 chains of samples initiated with different values of parameters. We begin by observing marginalised 1-d projections at $95\%$ confidence level~(see Fig.~\ref{single_95}) obtained using MCMC at different luminosities and systematics. 
			We observed~(top row of Fig.~\ref{single_95}) that when systematics is chosen to be $(0,0)$, the confidence interval on anomalous couplings~($c_i$) tightens on increasing luminosity $\mathcal{L}$. The limits vary as $\approx \frac{1}{\sqrt{\mathcal{L}}}$. And when the systematics is chosen at $(0.25\%,0.5\%)$ for asymmetries and cross-section respectively, we note~(middle row of Fig.~\ref{single_95}) the confidence interval tightens with increasing luminosity $\mathcal{L} \in \left[100~\text{fb}^{-1},1000~\text{fb}^{-1}\right]$ and the limits on every anomalous couplings saturates above $1000~\text{fb}^{-1}$. The bottom row of Fig.~\ref{single_95} represents the variations of confidence interval of various anomalous couplings when the systematics are kept at $(1\%,2\%)$ and it is observed that the limits saturates at luminosity $\mathcal{L} = 250~\text{fb}^{-1}$.
            Similar arguments can be made by observing Fig.~\ref{Most corr pairs}, which is a marginalized 2-d projection of the correlated parameters at $95\%$ confidence level. Other combinations of parameters are found to be minimally correlated or uncorrelated. The plots are shown for three different set of systematic error at different luminosities. The left, middle and right panel of each row of Fig.~\ref{Most corr pairs} represents when systematics are chosen to be $(0,0)$,$(0.25\%,0.5\%)$ and $(1\%,2\%)$ respectively. We observed that as one moves from left panel to right panel~(increasing systematic error) in each row, the simultaneous limits on two anomalous couplings get saturated at certain luminosity. As can be noted from middle and right panel of Fig.~\ref{Most corr pairs}, the simultaneous limits on two parameters saturates around $1000~\text{fb}^{-1}$ and $250~\text{fb}^{-1}$ respectively. 
            Finally, we show in Fig.~\ref{lumvseps} the variation of limits of all five anomalous couplings w.r.t $\mathcal{L}$ at fix $\epsilon_{A}$ and $\epsilon_{\sigma}$. It depicts how the limits of different couplings varies w.r.t luminosities at a fixed systematic error. As can be seen in Fig~\ref{lumvseps}, the limits at systematics of $(1\%,2\%)$ for asymmetries and cross-section respectively shown by black curve saturates suggesting that the best limits set on various couplings does not improve with luminosity. It can be further noted that the limit set at chosen maximal luminosity~($3000~\text{fb}^{-1}$) and maximal systematics~($1\%,2\%$) is stil worse than when we chose minimal luminosity~($100~\text{fb}^{-1}$) but reduced systematics~($0.25\%,0.5\%$).
			For example, we take $c_{WWW}$, the limit at $(\epsilon_{A},\epsilon_{\sigma}) = (1\%,2\%)$ and $\mathcal{L}=100$ fb$^{-1}$ is approx. $[-2.2,+2.2]$ and these limit becomes approx. $[-2.0,+2.0]$ at a luminosity value of 3000 fb$^{-1}$. The later value can be achieved if we chose the systematics to be $(0.25\%,0.5\%)$ and luminosity is kept at 100 fb$^{-1}$. The same behaviour is shown by all the other anomalous couplings. These do suggest that unless the systematic error is not brought down to a sizeable value, increasing the luminosity would not be very advantageous. The limit for different parameters~($c_i$) at $95\%$ confidence interval for $(\epsilon_{A},\epsilon_{\sigma}) = (1\%,2\%)$ are noted down in Table~\ref{95bci}. We found that the limits set by our analyses on the parameters like $c_B$ and $c_{\widetilde{W}}$ is more tighter than that of the quoted experimental value in Table~\ref{tab:constraint} while the limits on other anomalous couplings~($c_{WWW},c_W,c_{\widetilde{WWW}}$) remains comparable. The limits on $c_{\widetilde{W}}$ improves by a factor of $\approx$ 1.18 and for $c_B$ the limit shrinks by a factor of $\approx$ 1.7. We further note that that the limit obtained in this article for anomalous couplings like $c_{W}$ and $c_B$ are tighter by a factor of $\approx 1.5$ and $2.3$ respectively then those obtained by~\cite{Rahaman:2019mnz}. The limits by~\cite{Rahaman:2019mnz} were obtained at $\sqrt{s}=500$ GeV at the level of parton, i.e there were no error in the reconstruction of $W^+$ boson.
			\begin{table}[!htb]
				\caption{\label{95bci}The list of $95\%$ BCI of anomalous
					couplings (TeV$^{-2}$) of effective operators for $\sqrt{s} = 250$ GeV and
					$\mathcal{L} \in \{100~\text{fb}^{-1}$, $250~\text{fb}^{-1}$,
					$1000~\text{fb}^{-1}$, $3000~\text{fb}^{-1}\}$ at sytematic error of $(\epsilon_{A}, \epsilon_{\sigma}) = (1\%,2\%)$ from MCMC global fits at the
					\emph{reconstruction level}. The reconstruction of $W^+$ is done using {\tt XGBoost}.}
				\centering
				\begin{ruledtabular}
					\begin{tabular}{ccccc} 
					  Parameters&$100$ fb$^{-1}$&$250$ fb$^{-1}$&$1000$ fb$^{-1}$&$3000$ fb$^{-1}$ \\ \hline \\[-1em]
					  $\frac{c_{WWW}}{\Lambda^2}$&$^{+2.1}_{-2.3}$&$^{+1.8}_{-1.9}$&$^{+1.6}_{-1.7}$&$^{+1.5}_{-1.5}$\\ \\[-0.9em]
					  $\frac{c_{W}}{\Lambda^2}$&$^{+2.3}_{-2.0}$&$^{+1.9}_{-1.7}$&$^{+1.6}_{-1.5}$&$^{+1.6}_{-1.4}$\\ \\[-0.9em]
					  $\frac{c_{B}}{\Lambda^2}$&$^{+5.0}_{-6.4}$&$^{+4.2}_{-5.3}$&$^{+3.6}_{-4.5}$&$^{+3.5}_{-4.3}$\\ \\[-0.9em]
					  $\frac{c_{\tilde{W}}}{\Lambda^2}$&$^{+17.7}_{-17.8}$&$^{+15.3}_{-15.4}$&$^{+13.4}_{-13.5}$&$^{+12.8}_{-13.0}$\\  \\[-0.9em]
					  $\frac{c_{\widetilde{WWW}}}{\Lambda^2}$&$^{+2.6}_{-2.6}$&$^{+2.1}_{-2.1}$&$^{+1.8}_{-1.8}$&$^{+1.7}_{-1.7}$\\ 
					\end{tabular}
				\end{ruledtabular}
			\end{table}\\ 
\section{Conclusion}
\label{conclude}
In this article, we use machine learning techniques like
artificial neural network and boosted decision trees to tag the jets initiated
by light flavor quarks with about $70\%$ accuracy. The classification is made on
two class defined as up-type or down-type jets. 
The reconstructed $W^+$ still remains a faithful object for constructing all the
observables that we have used in this work to put a constraint on the anomalous
couplings. It is always an advantage to have as many observables as possible to
probe or measure various parameters of new physics and we have shown that
spin-spin correlation asymmetries can provide constraints that are comparable to
the constraint provided by polarization asymmetries. We also exploit the fact
that the spin and polarization asymmetries vary with the production angle
$\theta_{W^-}$ by dividing the $\cos\theta_{W^-}$ in eight equal bins and
constructing all 80 asymmetreis in all those bins. This along with the
cross-sections in those bins gives us a total of 648 observables that depende
upon the five anomalous couplings.

Our one parameter limits shown in Table~\ref{onepara95ci} for ${\cal L}=100$ fb$^{-1}$ are
better than the one parameter limits in Table~\ref{tab:constraint} obtained from various
analyses~\cite{CMS:2021foa,CMS:2021icx,CMS:2019ppl} at CMS.
Our five parameter simultaneous limits in Table~\ref{95bci} for ${\cal L}=100$ fb$^{-1}$
are also comparable to the one parameter limits in Table~\ref{tab:constraint} for $c_{W}$, $c_{B}$
and $c_{\tilde{W}}$. While for $c_{WWW}$ and $c_{\widetilde{WWW}}$ the CMS
limits obtained using production rates alone are better than what we quote. This
is because the contributions proportional to $c_{WWW}$ and
$c_{\widetilde{WWW}}$ have extra factor of $p^2$ in the matrix-element, which
leads to an enhanced contribution in machine like LHC running at 13 TeV. In our
case the limit $c_{\widetilde{WWW}}$ is derived mainly from the asymmetries and
without the advantage of large momentum. For $c_{WWW}$, however, the
cross-section provides strong limit Fig.~\ref{1dplot} but there is some cancellation in
the cross-section due to non-zero values of $c_{W}$ and $c_{B}$ Fig.~\ref{xgbcwwwcwcwcwt} which
leads to a poorer limit than Table~\ref{onepara95ci} when all parameters are varied.

We showed how systematic error act as a brick on constraining anomalous
couplings. For example, for a conservative choice of systematics to be
($\epsilon_{A},\epsilon_{\sigma}$)=($1\%,2\%$), the limits on the anomalous
coupligs improve by a factor of only $\sim1.4$ when we increase the luminosity
from 100 fb$^{-1}$ to 3000 fb$^{-1}$. This indicates that with large systematics
it is wise to look for additional observables from various processes to better
constrain the couplings than to run the machine for a higher luminisity. One can
also try to improve the flavor tagging, use the beam polarization, use a finer
binning of the production angle $\cos\theta_{W^-}$ while running the machine at
$\sqrt{s} = 250$ GeV.

			
	\begin{acknowledgments}
		We thank Rafiqul Rahaman for useful discussions. A.S thanks CSIR-UGC, Government of India for financial support. 
	\end{acknowledgments}
	
	\appendix
	\section{Normalised polarisation and decay density matrix}
	The polarisation density matrix for a spin-1 particle is given as $\rho(\lambda,\lambda') = $
	\begin{equation}
		\label{eqn:A1}
		\begin{bmatrix}
			\frac{1}{3}+\frac{p_z}{2}+\frac{T_{zz}}{\sqrt{6}}&\frac{p_x-ip_y}{2\sqrt{2}}+\frac{T_{xz}-iT_{yz}}{\sqrt{3}}&\frac{T_{xz}-T_{yy}-iT_{xy}}{\sqrt{6}} \\
			\frac{p_x+ip_y}{2\sqrt{2}}+\frac{T_{xz}+iT_{yz}}{\sqrt{3}}&\frac{1}{3}-\frac{2T_{zz}}{\sqrt{6}}&\frac{p_x-ip_y}{2\sqrt{2}}-\frac{T_{xz}-iT_{yz}}{\sqrt{3}}\\
			\frac{T_{xx}-T_{yy}+2iT_{xy}}{\sqrt{6}}&\frac{p_x+ip_y}{2\sqrt{2}}-\frac{T_{xz}+iT_{yz}}{\sqrt{3}}&\frac{1}{3}-\frac{p_z}{2}+\frac{T_{zz}}{\sqrt{6}}
		\end{bmatrix}
	\end{equation}
	and the normalised decay density matrix $\Gamma(\lambda,\lambda') = $
	\begin{equation}
		\label{eqn:A2}
		\begin{bmatrix}
			\frac{1+\delta+\rho c^2_\theta+2\alpha s_\theta}{4}&\frac{s_\theta+d\delta c_\theta}{2\sqrt{2}}e^{i\phi}&\rho \frac{(1-c^2_\theta)}{4}e^{i2\phi}\\		
			\frac{s_\theta(\alpha + \rho c_\theta)}{2\sqrt{2}}e^{-i\phi}&\delta+\rho \frac{s_\theta^2}{2}&\frac{s_\theta(\alpha-\rho c_\theta)}{2\sqrt{2}}e^{i\phi} \\
			\rho \frac{(1-c^2_\theta)}{4}e^{-i2\phi}&\frac{s_\theta(\alpha-\rho c_\theta)}{2\sqrt{2}}e^{-i\phi}&\frac{1+\delta+\rho c_\theta^2 - 2\alpha c_\theta}{4}
		\end{bmatrix}
	\end{equation}
	where $c_\theta$ and $s_\theta$ are cos$\theta$ and sin$\theta$ of the polar angle of the decay products respectively and $\rho = (1-3\delta)$.
			\bibliography{biblio}
		\end{document}